\documentclass{article}
\usepackage[a4paper, total={6.5in,9in}]{geometry}
\usepackage{graphicx}
\usepackage{epstopdf}
\usepackage{subfigure}
\usepackage{url}
\usepackage{authblk}
\usepackage{amsmath}
\usepackage{amsfonts}
\usepackage{natbib}

\begin{document}


\title{Financial ratios and stock returns reappraised through a topological data analysis lens}

\vspace{-30pt}
\author[1]{Pawe{\l} D{\l}otko\thanks{Full Address: Mathematics Department, College of Science, Swansea University, Bay Campus, Swansea, SA1 8EN, United Kingdom. Email:p.t.dlotko@swansea.ac.uk. }}
\affil[1]{Mathematics Department, Swansea University, United Kingdom}
\vspace{-40pt}
\author[2]{Wanling Qiu\thanks{Full Address: Accounting and Finance Subject Group, School of Management, University of Liverpool, 20 Chatham Street, Liverpool, L69 7ZH, United Kingdom. Email:wanling.qiu@liverpool.ac.uk}}
\affil[2]{School of Management, University of Liverpool, United Kingdom}

\vspace{-30pt}

\author[3]{Simon Rudkin \thanks{\textbf{Corresponding Author}. Full Address: Economics Department, School of Management, Swansea University, Bay Campus, Swansea, SA1 8EN, United Kingdom. Tel: +44 (0)1792 606325 Email:s.t.rudkin@swansea.ac.uk}}
\affil[3]{Economics Department, Swansea University, United Kingdom}
\vspace{-30pt}

\vspace{-20pt}

\maketitle

\begin{abstract}
Firm financials are well established as return predictors, being the inspiration for a large set of anomalies in the asset pricing literature. Employing topological data analysis we revisit the question of association between seven of the most commonly studied financial ratios and stock returns. Specifically the TDA Ball Mapper algorithm is applied to visualise the point cloud of financial ratios as an abstract two-dimensional graph readily allowing for identification of interdependencies between factors. These relationships are seldom monotonic, opportunities for investors to profitably exploit this knowledge provided by TDA abound. Clear potential offered by the tools of TDA to shed new light on asset pricing models is demonstrated. Scope for benefit is limited only by the availability of information to the analyst.
\end{abstract}


\section{Introduction}
\label{sec:intro}

Stock returns are intrinsically linked to firm financial ratios with a vast literature seeking the optimal model of exactly how \citep[for example]{green2017characteristics}. Data is abundant and the toolkit with which the relationship is modelled is ever expanding. However, the principles upon which stock return models are constructed remain broadly the same as those that drove the seminal work of \cite{fama1993common}; potential anomalies are identified, sorts performed and a potential new factor produced. Consequently a ``factor zoo'' has emerged but in every case inclusion is premised on a monotonic relationship \citep{stambaugh2016mispricing,feng2017taming,fama2018choosing}. This paper casts new light on the process of anomaly identification, illuminating several key departures from monotonicity. It achieves this aim through the application of Topological Data Analysis (TDA), an approach widely adopted in the physical sciences which has much to offer to the study of stock returns. 

TDA, formalised in \cite{carlsson2009topology}, considers multivariate datasets as point clouds and seeks features therein. A point cloud is simply a scatter plot taken to multiple dimensions, more data facilitating the inclusion of more axes to define the space. In traditional statistical analysis an underlying distribution is assumed from which data is drawn randomly. TDA, on the other hand, builds up descriptors of point clouds without any assumptions on the underlying distribution. In addition, these descriptors are robust to noise as well as continuous transformations of the point cloud. This paper will build two-dimensional models of seven-dimensional data sets, but the principles discussed extend readily to point clouds with as many axes as there are potential anomalies identifiable. 

To motivate the application herein, consider a collection of measurements sampled from the surface of the Earth, and TDA as a mapping tool. Potential areas with no
information are holes in the knowledge and will be identified by the TDA.
Subsequently, proximity and relationships within the point cloud can be used to extrapolate the information into the uncharted regions and therefore get the complete picture, or a map. That allows inference of a continuous picture from a finite collection of observables, without imposing any model of the data. This is very much like a cartographer plots a continuous map based on a finite, incomplete information set.
In this context monotonic relationships between two mis-pricing factors and the stock returns will produce a high dimensional map that rises (or falls) along the length of each axis. For the bivariate case of size and book-to-market \citep{fama1993common}, the highest part of the surface would match small firms with high book-to-market ratios. 

Demonstration of the ability of TDA Ball Mapper to inform on stock returns requires selection of a subset of common anomalies whose effects are well established in the literature. To this end a natural starting point is the five factor model of \cite{fama2015five} (FF5) which augments the book-to-market and size factors from \cite{fama1993common} (FF3) with profitability and investment factors. This extends the number of axes to four since the fifth factor, the market return, only takes a single value for all firms in a month. To this set, three further firm financial ratios are added, being the earnings-to-price ratio, the cashflow-to-price ratio and the dividend yield. In total a set of seven common anomalies is thus used in the analysis that follows.

In the original capital asset pricing model (CAPM) of \cite{lintner1965security,sharpe1964capital,treynor1962jack} the only explanation of stock returns came in a risk free return, the ``alpha'', and the slope coefficient relative to the market, the ``beta''. However, it was quickly noted that there were many other variables that have statistical significance in the prediction of stock returns. \cite{ou1989financial} work considered a number of financial ratios from firms accounting data to find significant links to the stock returns achieved. These results gathered criticism because they were driven by data rather than the theory of why each variable might be important \cite{fama1992cross}. Thirty years later there is still a wealth of literature which takes this data driven approach where new variables are found in an agnostic way, and subsequently appropriate theoretical explanations are found. 

Of the variables that emerged in this early period of exploration the dividend yield \citep{black1974effects,fama1988dividend}, earnings-to-price \citep{basu1977investment,ball1992}, and cashflow-to-price \citep{haugen1996commonality}  all continue to find theoretical and empirical support in contemporary studies. Indeed these are the variables offered by Ken French as potential sorts for model benchmarking on his well cited data website\footnote{Sorted portfolio returns based on these, and the FF5 factors can be downloaded from \url{https://mba.tuck.dartmouth.edu/pages/faculty/ken.french/data_library.html}.}. Many others have been dismissed as being spanned by either these three, or the FF5 factors. These three, plus the FF5 thus go forward to the subsequent analysis.


The most common way of identifying the new firm financial variables $x$ as a determinant of stock returns is by using a linear regression of returns on $x$ together with the appropriate controls that are already known to influence the returns. If the coefficient of the new variable $x$ is significant then $x$ is a candidate for being a new anomaly. Incorporation of the new $x$ in a new model can then either be done through using $x$ directly in the model as in \cite{ou1989financial}, or through the use of sorted portfolio returns, sorting on $x$, as in \cite{fama1993common} and similar. This approach suffers from a significant handicap: it assumes linear dependence of returns from data as well as the relation from individual variables rather then their combinations. This paper evidences that such linearities are not as common as the modelling strategy of factor studies would have the reader believe. With no requirement on functional form to map characteristics to outcome, this inability to perform a meaningful regression is not of particular concern to TDA analysis. Subsequent results demonstrate that moving away from assumptions of linear relationships is advisable.


As presented there is an impression that TDA Ball Mapper is producing a clustering of points of the point cloud, and that it subsequently seeks to map the clusters to the returns. However, such a two-stage cluster then consider returns approach has already been used. \cite{liao2008mining} being one of very few examples to have first clustered on firm characteristics. Such approaches have not gained much traction in the literature. The main reason for low adoption is because clusters coming from a typical clustering algorithm seldom provide information about the links from the right hand side variables, the explanatory factors, to the outcome variable. Those clusters are chosen for objectives such as minimising average variation used in k-means clustering \citep{hartigan1979algorithm}. Hence clusters may have considerable size and complicated shapes; the distances between cluster centroids typically vary widely. On the contrary to that, the TDA Ball Mapper is based on covering the point cloud with balls of the same radii. That gives a sharp bound on the possible distance between connected points; topographical detailing is preserved. Hence, when trying to compare the returns on groups of TDA Ball Mapper balls it is much easier to make inference on the outcome variable. More evidence on the key differences between TDA Ball Mapper and clustering follows through theory and empirics.

TDA Ball Mapper is not limited in the number of dimensions that can be represented in a standard plot. In that regard, ready extension may be made to keep pace with the ever growing literature on the ``factor zoo''.  Large datasets with hundreds of financial ratios are already being exploited by data scientists \citep[for example]{yan2017fundamental}  as well as those using more traditional regression approaches \citep{green2017characteristics,stambaugh2016mispricing}. There are a wealth of working papers that continue to explore how machine learning can create models from within the set, \cite{kozak2018interpreting} being one such example recently published. All that is discussed here may be applied to any of the datasets assembled for these machine learning papers. Where those works set out metrics to capture improved fit, and learn from the full possible model set, a TDA approach could quickly identify whether candidate anomalies improved fit. Using the abstract plot, and having the regression residual from the extended model as an outcome variable, it would be straightforward to see exactly where residuals were falling. Such represents a potential start point for TDA as an anomaly detection tool. There are many directions to expand there and such are beyond the scope of this paper.    

In two-dimensions visualisation of such surfaces is straightforward. However, given the early identification of multiple anomalies, and the growth of big data sources, there are far more dimensions than we could possibly map. Consequentially there is a call for a simpler representation that maintains the information content of all dimensions but can be seen in two-dimensions. The call is answered with the mapper algorithm of \cite{dlotko2019ball}. Primary contributions of this paper are thus a new understanding of the links between established stock return factors and returns, removal of the reliance on pre-assumed relationships within factors, and the creation of a roadmap for future research on the topological analysis of stock return data. From a practitioner perspective this paper presents the first evidence that there are unexploited properties of existing easily accessible data that have strong links to trading profit.

The remainder of the paper is organised as follows. Before outlining the method, data for this study is introduced in Section \ref{sec:data}. A theoretic and context relevant exposition of the TDA Ball Mapper algorithm is provided in Section \ref{sec:tda}. Section \ref{sec:bi} develops the interpretation of TDA Ball Mapper output through a simple two dimension example after FF3. Extending the set to the full seven common anomalies, Section \ref{sec:sort} shows where in the parameter space models are performing best. With monthly data it is natural to study monthly point clouds, but TDA Ball Mapper can be applied to much larger datasets as Section \ref{sec:annual} demonstrates. Section \ref{sec:discuss} reviews further oppotunities for analysis and extends consideration of the next steps. Finally, Section \ref{sec:conclude} concludes.

\section{Data}
\label{sec:data}

Taking stock data from the Center for Research in Stock Prices (CRSP) and Compustat, firms' financial ratios are constructed using code from \cite{green2017characteristics}. This paper considers data relating to stock returns between July 1976 and December 2018, but only a section of this is used in the illustrations. In the examples that follow five year-month combinations are used, being June 1978, June 1988, June 1998, June 2008 and June 2018. These specific months are motivated as ten year intervals either side of the turning point in US GDP in the global financial crisis. In what follows these ten year intervals are used for illustration, but results are robust to the choice of other months. Table \ref{tab:sumstats} provides summary statistics for the whole dataset as well as for June 2018. 

\begin{table}
 \begin{center}
     \caption{Summary Statistics}
    \label{tab:sumstats}
    \begin{tiny}
    \begin{tabular}{l c c c c c c c c}
         \hline
         Variable & Mean & s.d. & Min & q25 & q50 & q75 & Max\\
         \hline
         \multicolumn{8}{l}{Panel (a): Full Sample ($n=1863690$)} \\
         Returns (\%) & 0.619 & 14.74 & -76.09 & -6.663 & 6.787 & 266.1\\
         Size (log market value) & 5.008 & 2.189 & -1.135 & 3.398 & 4.888 & 6.533 & 12.36 \\
         Book-to-market & 0.728 & 0.749 & -62.37 & 0.330 & 0.588 & 0.955 & 30.00 \\
         Profitability (return-on-equity) & 0.018 & 0.533 & -12.23 & -0.015 & 0.099 & 0.177 & 18.22 \\
         Investment & 0.076 & 0.170 & -0.932 & 0.001 & 0.039 & 0.113 & 2.658 \\
         Earnings-to-price & -0.018 & 0.324 & -11.99 & -0.008 & 0.048 & 0.084 & 0.614 \\
         Cashflow-to-price & -1.592 & 52.70 & -1063 & -8.898 & -0.568 & 4.704 & 1441 \\
         Dividend yield (\%) & 1.598 & 2.649 & 0 & 0 & 0.018 & 2.443 & 63.95 \\
         \multicolumn{8}{l}{Panel (b): June 2018 ($n=2837$)}\\
         Returns (\%) & 0.350 & 10.02 & -40.96 & -4.328 & 0.097 & 4.799 & 52.17\\
         Size (log market value) & 6.747 & 2.091 & 1.786 &5.244 & 6.797 & 8.159 & 12.290 \\
         Book-to-market & 0.498 & 0.636 & -5.120 & 0.225 & 0.417 & 0.669 & 16.27 \\
         Profitability  (return-on-equity) & -0.049 & 0.813 & -7.249 & -0.103 & 0.072 & 0.142 & 10.86\\
         Investment & 0.026 & 0.080 & -0.372 & -0.001 & 0.010 & 0.048 & 0.656 \\
         Earnings-to-price & -0.057 & 0.277 & -3.127 & -0.042 & 0.031 & 0.052 & 0.319 \\
         Cashflow-to-price & 2.210 & 35.02 & -249.6 & -3.434 & 1.271 & 7.093 & 348.5\\
         Dividend yield (\%) & 1.234 & 2.169 & 0 & 0 &0.015 & 1.831 & 22.33\\
         \hline
    \end{tabular}
    \end{tiny}
 \end{center}
\raggedright
\footnotesize{Notes: Data sourced from CRSP and Compustat using code from \cite{green2017characteristics}. All data is winzorised monthly at the 1\% level to limit the impact of outliers.}
\end{table}

Construction of the variables follows \cite{green2017characteristics}. Compustat variable names are given in parentheses. Size is equal to the log of market value (\textit{mvef}) and book-to market ratio is the book value of the firm's equity divided by the market value ($ceq/mvef$). Profitability is measured as the income before tax divided by the lagged value of equity ($ib/lag(ceq)$). Investment is calculated as the sum of increases in property plant and equipment ($ppegt$) and inventory ($invt$) divided by the lag value of total assets ($ppegt-lag(ppegt)+invt-lag(invt))/lag(at)$. Earnings-to-price ratio is simply the income before tax divided by the market value  ($ib/mvef$), cashflow to price is the sum of long term debt ($dltt$) and market value ($mvef$) less the total value of the firm's assets ($at$) divided by cash and short term equivalents ($che$), that is $(dltt+mvef-at)/che$. Finally the dividend yield is taken directly from CRSP.  

For all variables summarised in Table \ref{tab:sumstats} the interquartile range is small relative to the overall range. It is later shown how this creates a common mass at the centre of the TDA Ball Mapper plots. Because of the variation in the levels of the seven common anomalies, normalisation is applied to place each on the scale $[0,1]$. All variables are also winzorised on a monthly basis to remove the top, and bottom, 0.5\%. These reduced monthly samples are then combined to create samples for longer time frames. 

\section{TDA Ball Mapper}
\label{sec:tda}

Understanding high dimensional and complicated data is a keen pursuit for the machine learning and data science community.  The input to the presented analysis is a $d$ dimensional point cloud $X$, i.e. any given datapoint is defined by its coordinates on the $d$ axes $(x_1,x_2,...,x_d)$. In addition, for every pair of points $x,y \in X$ the metric of similarity measure $dist(x,y)$ is given.

In this paper we use mathematically rigorous tools coming from \emph{Topological Data Analysis} (TDA), in particular the \emph{Ball Mapper algorithm} developed by \cite{dlotko2019ball}. In what follows the algorithm is given a fuller title, TDA Ball Mapper, to distinguish it more clearly from older mapper algorithms. TDA Ball Mapper is inspired by the \emph{Reeb} and \emph{Conventional Mapper graphs}\footnote{See \cite{carriere2018structure} for a further review of the original mapper software.}. Prior to exposition of the \cite{dlotko2019ball} TDA Ball Mapper algorithm an overview of the Conventional Ball Mapper is given. 

\subsection{Conventional Ball Mapper}

The \emph{Conventional Mapper algorithm} offers a way of encapsulating the shape of a point cloud $X$, equipped with a function $f : X \rightarrow \mathbb{R}^n$ (typically $n=1$). To achieve it, a \emph{cover} of the range of $f$ with a collection of overlapping sets (typically intervals) $I_1,\ldots,I_n$ is constructed. The Conventional Mapper graph is an abstract graph $G$ having vertices and edges defined as follows. For every interval $I_k$, the clusters in $f^{-1}(I_k)$ are considered. Each cluster corresponds to a vertex in $G$. Two clusters $C_1$ and $C_2$, with a nonempty intersection, give rise to an edge between the vertices of $G$ corresponding to $C_1$ and $C_2$. Note that, if only hard clustering is used, there can be nonempty intersections only for clusters in inverse images of the distinct elements $I_k$, $I_{l}$ covering $\mathbb{R}^n$ such that $I_k \cap I_{l} \neq \emptyset$.

As one can deduce from the description, there are a number of parameters that need to be set up to obtain the Conventional Mapper graph. Those are the function $f$, coverage of $\mathbb{R}^n$ and the clustering algorithm used to obtain clusters in $f^{-1}(I_k)$. The main idea of the Conventional Mapper algorithm aims to obtain certain coverage of points in $X$. The TDA Ball Mapper Algorithm presented in the next section does likewise, but is able to build such a cover of $X$ in a more direct and geometrically faithful way.

\subsection{Ball Mapper Algorithm}
The main purpose of the TDA Ball Mapper algorithm is to create a cover of the space $X$ with a collection of overlapping sets $C_1,\ldots,C_n$, for a certain $n>1$, each of which is a subset of $X$. The number of sets, $n$, is determined as an outcome of the algorithm as described below. Moreover it is assumed that the points gathered in each $C_i$ are
geometrically close (this vague requirement will also be specified in detail later on). Given the collection $C_1,\ldots,C_n$, the vertices of the Ball Mapper Graph correspond to the cover elements and, for simplicity, further assume that the vertex $i$ corresponds to $C_i$. The vertices $i,j$ in $G$ are joined by 
an edge if and only if $C_i \cap C_j \neq \emptyset$, i.e. the corresponding clusters have nonempty
intersection. Clearly in this case one can define \emph{weights} of both vertices and edges of $G$. In our
case, we will use the cardinality (i.e. the number of elements) of $C_i$ as a weight of the vertex $i$ and the number of elements in $C_i \cap C_j$ (if nonzero) as a weight of an edge between $i$ and $j$. They will be typically visualised by varying size of vertices and edges in the plot of the graph.

It remains to explain how the cover $C_1,\ldots,C_n$ is obtained. There are a number of ways to achieve it and in what follows focus is placed on the most intuitive one. This method for obtaining a cover is based on the concept of \emph{$\epsilon$-net}. Consider a collection of points $X$ equipped with a distance or similarity measure $dist : X \times X \rightarrow \mathbb{R}$. Fix a number $\epsilon > 0$. An $\epsilon$-net is a subset of points $X' \subset X$ such that for every $x \in X$ there exist $x' \in X'$ such that $dist(x,x') \leq \epsilon$. There exist a number of ways of constructing such subsets and this paper will describe the most readily interpreable one. The first point $a$ of $X'$ is taken at random from $X$. Subsequently all the points in $X$ that are not farther away than $\epsilon$ from $a$ are marked as \emph{covered}. If there is still a point $b \in X$ that is not covered, it is added to $X'$ and the procedure is repeated until all points in $X$ are covered. Given this, the number $n$ of cover elements, while depending on $\epsilon$, is not given by any closed expression depending on $\epsilon$ and is therefore determined by the algorithm above. In the next subsection this process is illustrated for a two dimensional simple example. 

In the construction above all points in $X$ that are at most $\epsilon$ away from $a \in X'$, where $X'$ is an $\epsilon$-net, will form a single cover element. As they all come from a ball or a radius $\epsilon$, they are at most $2\epsilon$ away from each other, and therefore can be considered being \emph{geometrically close}. The name of the TDA Ball Mapper algorithm is accredited to the fact that cover of the space $X$ used herein is made by using balls. The radius $\epsilon$ is the only parameter required by the TDA Ball Mapper algorithm. One should think about it as a level of resolution one wants to use to examine data; small values will show detailed structures in $X$, while large ones will give the overall idea of the layout of $X$.

This construction gives a way of representing the shape of a complicated and high dimensional point cloud $X$ by using an abstract graph. However in addition thereto it often happens that the points of $X$ are accompanied by a function $f : X \rightarrow \mathbb{R}$ that is of interest. 
Now define the function $\hat{f}$ induced by $f$ on the Ball Mapper Graph $G$; $\hat{f} : G \rightarrow \mathbb{R}$. For every cover element $C_i$, the value $\hat{f}(i)$ is an average value of $f$ on elements of $C_i$. Note that for continuous $f$'s, its value restricted to $C_i$ will not vary much; $f$ will be close to $\hat{f}$. 

The function $\hat{f}$ on the TDA Ball Mapper graph is subsequently visualized by a proper colouration of the vertices of $G$. Doing so commends the TDA Ball Mapper Algorithm as an advanced way of plotting relations $(x,f(x))$, for which $x$ is sampled from a complicated and high dimensional set $X$. It often allows the location and understanding of various complicated relations between variables in $X$. This understanding is the primary motivation for the empirical examples that follow. 

A primary use for the TDA Ball Mapper graph is as a summary visualisation to guide further investigation. In this way the colouration function applied on the vertices is important. By observing variation in outcome across the cover, interesting cases may identify themselves. For example restricted areas where $f$ varies between its low and high values are obviously of interest as they identify combinations of the axis variables and recognise the joint effect of small variations amongst them. In such cases most monotonic functions would suggest that group should have similar $f$ values because of their similarities on all axes. This is seen in context in the applications. The user is cautioned to consult summary statistics for the balls to inform their conclusions.

In the context of this paper, interpretation is thus summarised as first an understanding of the shape, second an exploration of the contribution of the axes of $X$, and third an analysis of any seeming statistical anomalies that large variation in $f$ would represent within compact regions of $X$. Only one parameter is required, the radius $\epsilon$. Choice of $\epsilon$ is a trade off between reducing the number of balls to something manageable, seeking connectivity amongst the data, and ensuring that there are not important features getting subsumed within a higher level of aggregation. In this paper all examples are considered against this triple objective carefully, with robustness carried out for other radii as appropriate. 



 \subsection{Illustrative Example}
 
\begin{figure}
	\centering
	\caption{Development of Stylised Data Set.} \label{fig:mapeg1}
	\begin{tabular}{p{6cm} c}
		\includegraphics[height=6cm]{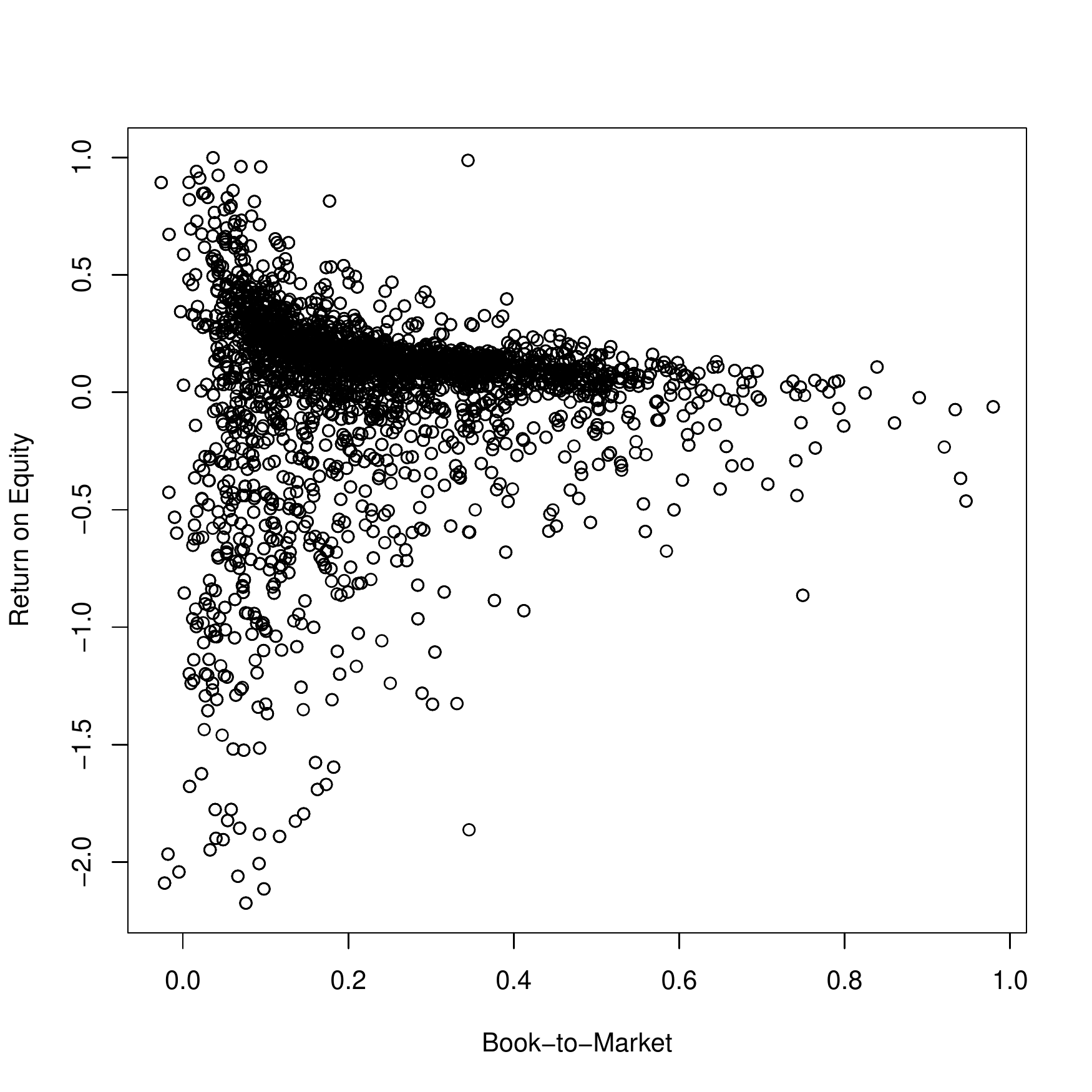}&
		\includegraphics[height=5.2cm]{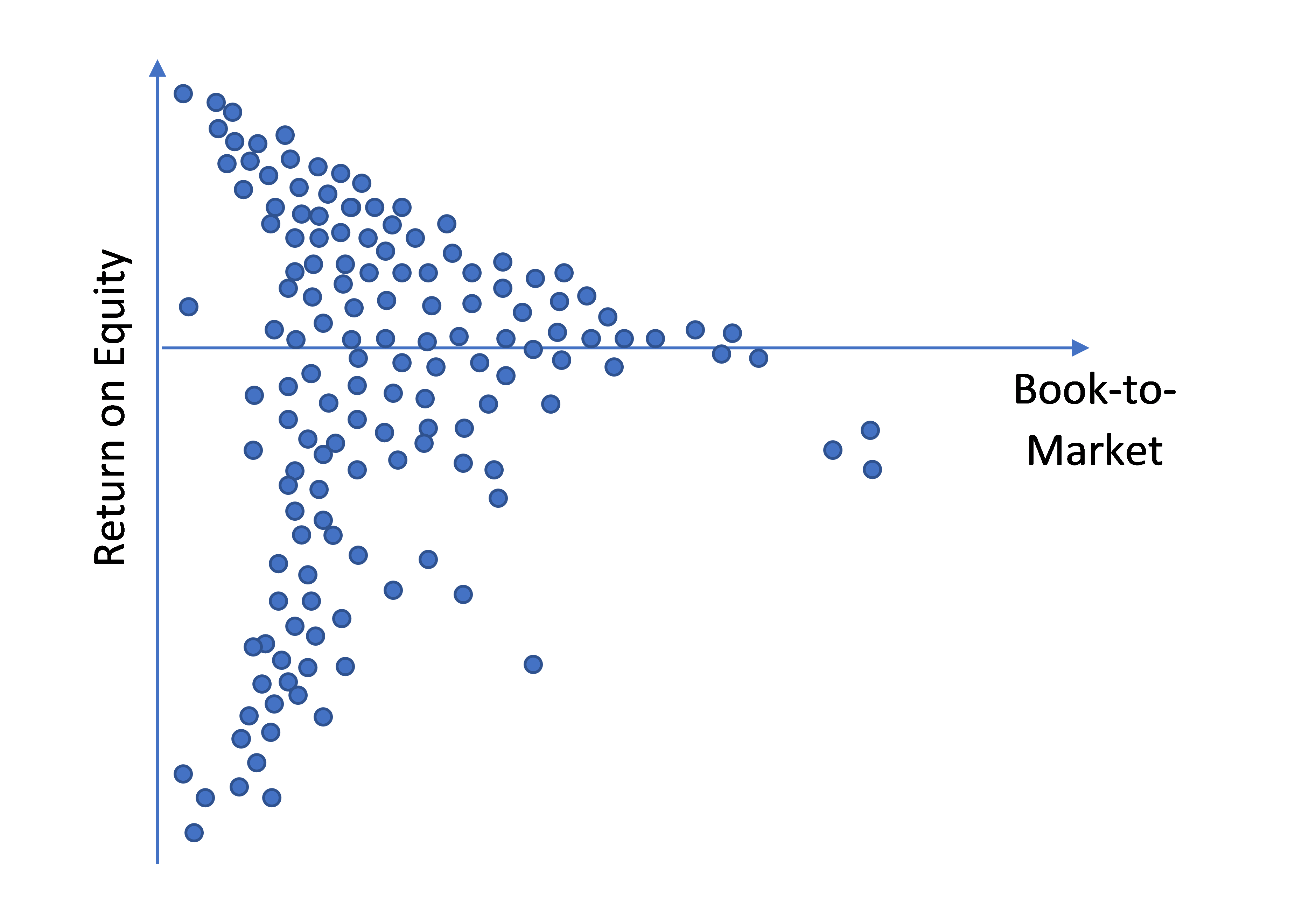}\\
		(a) 2016 book-to-market and return on equity & (b) Stylised plot \\
	\end{tabular}
\raggedright
\footnotesize{Notes: This plot uses data compiled following \cite{green2017characteristics} from CRSP and Compustat. Panel (a) data is for 2016. Panel (b) is a stylised representation.}
\end{figure}

 By way of an illustration of the TDA Ball Mapper algorithm consider just two variables from the set of seven common anomalies used in this paper. Book-to-market ratio and return on equity are employed for the demonstration, being distinct from those discussed in the subsequent results sections. Figure \ref{fig:mapeg1} presents the actual scatter plot, and a stylised version, constructed from 2016 data. There is a much larger range of return on equity associated with low book-to-market, and vice versa. Most of the data is in the range 0.1 to 0.4 on both axes but note that the scales on the two axes are different.

All that follows in the discussion of \ref{fig:mapeg1} is intuitive as humans readily understand two dimensional representations. This exposition is to intuit to show how real data may be converted to an abstract two-dimensional form. As noted in the methodological exposition, only one parameter is required for the process, the radius $\epsilon$. For any given $\epsilon$ a graph is formed that inherits the shape of the point cloud $X$.

\begin{figure}
	\begin{center}
	    \caption{Construction of the TDA Ball Mapper Graph} 
	\label{fig:mapeg2}
	\begin{tabular}{c c}
		\includegraphics[height=4.7cm]{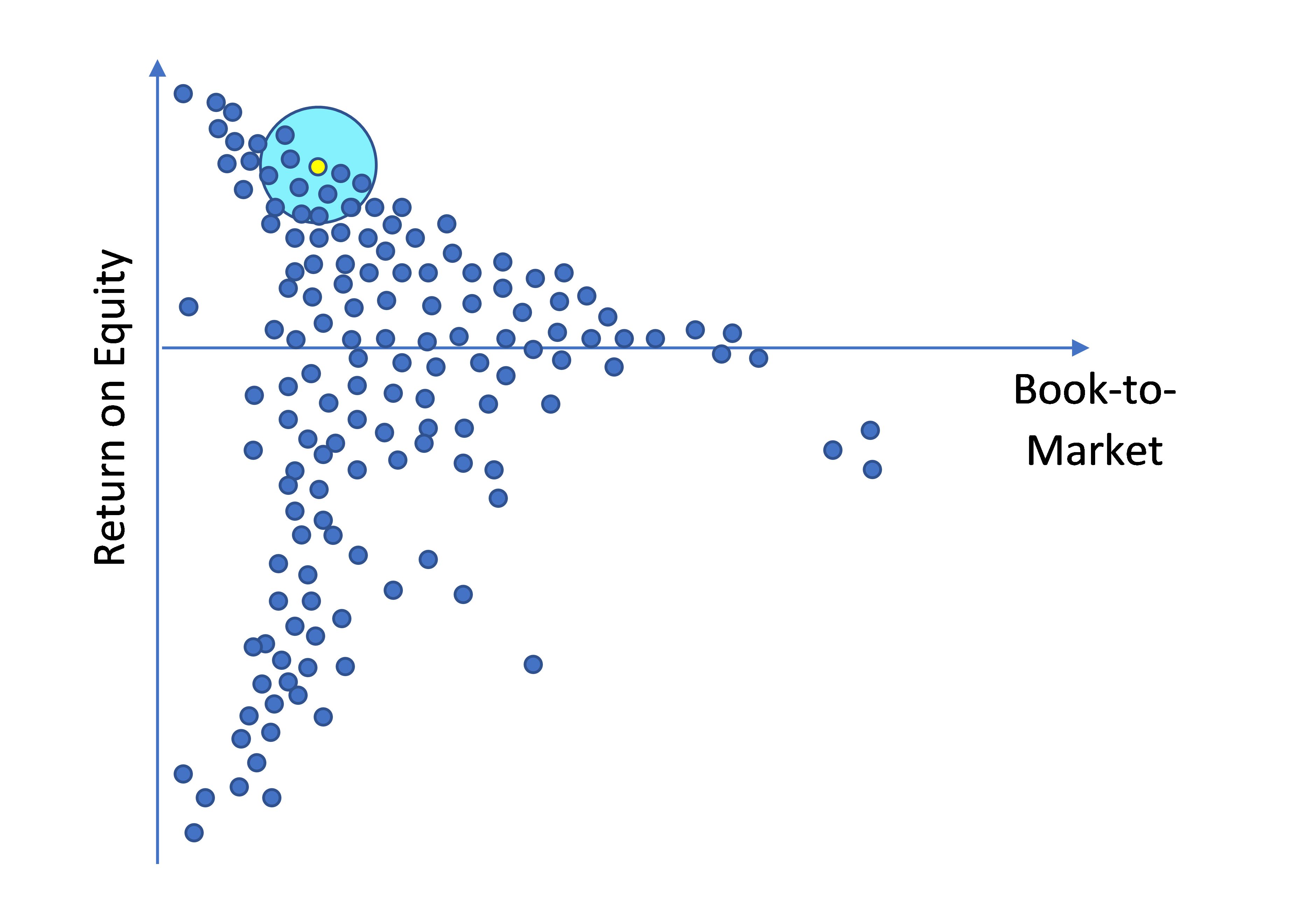}&
		\includegraphics[height=4.7cm]{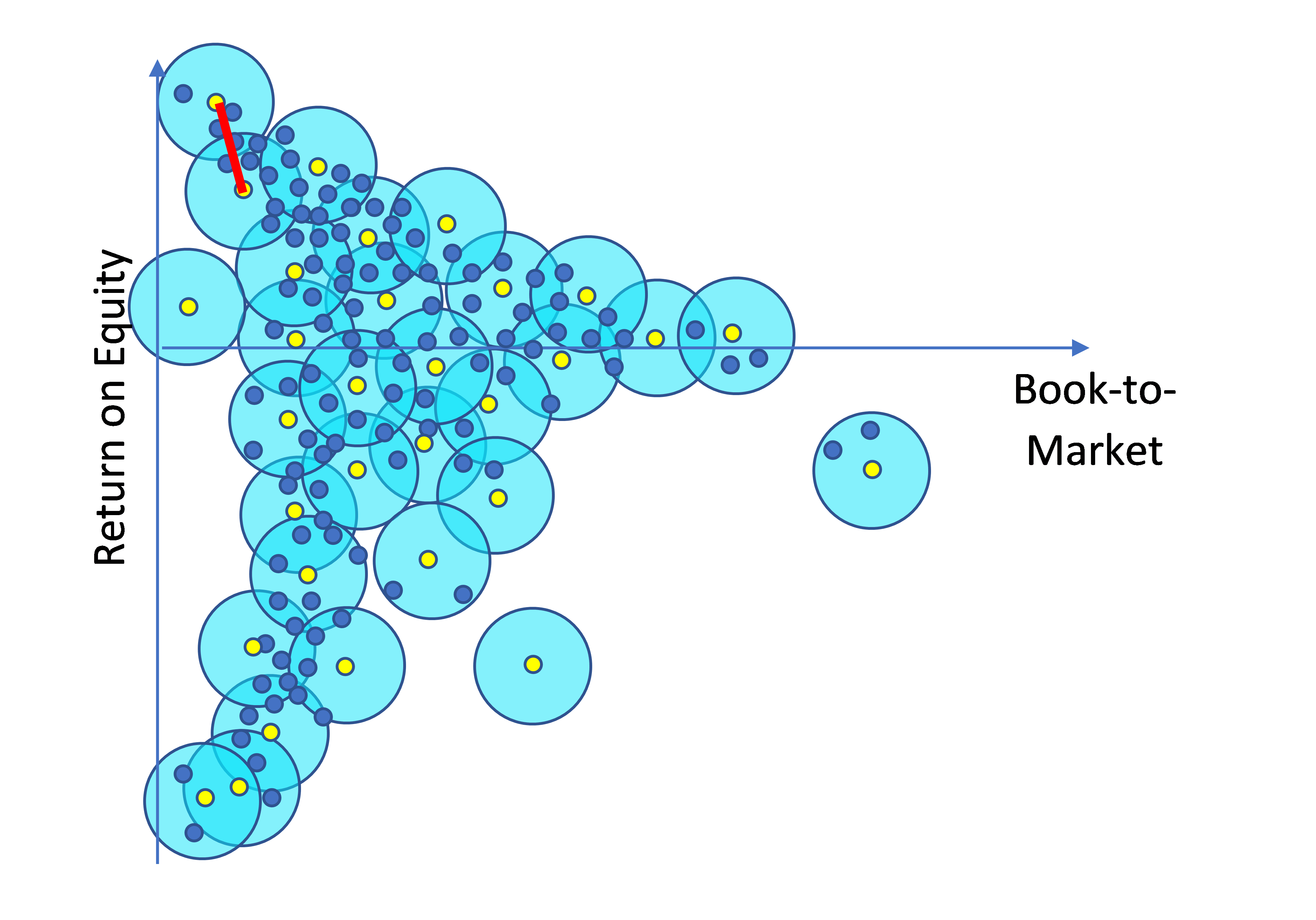}\\
		(a) An initial ball & (b) An initial edge \\
		\includegraphics[height=4.7cm]{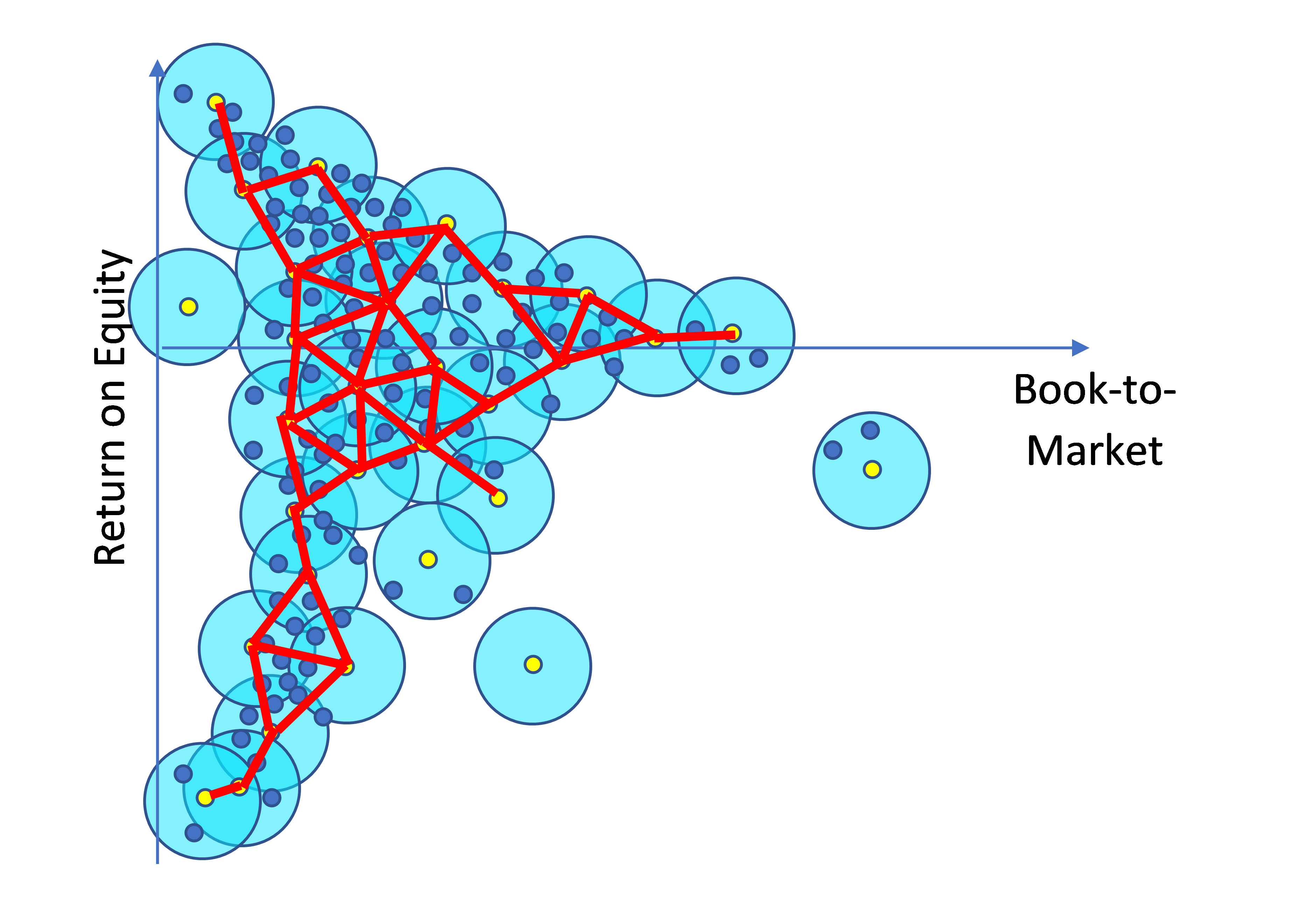}&
		\includegraphics[height=4.7cm]{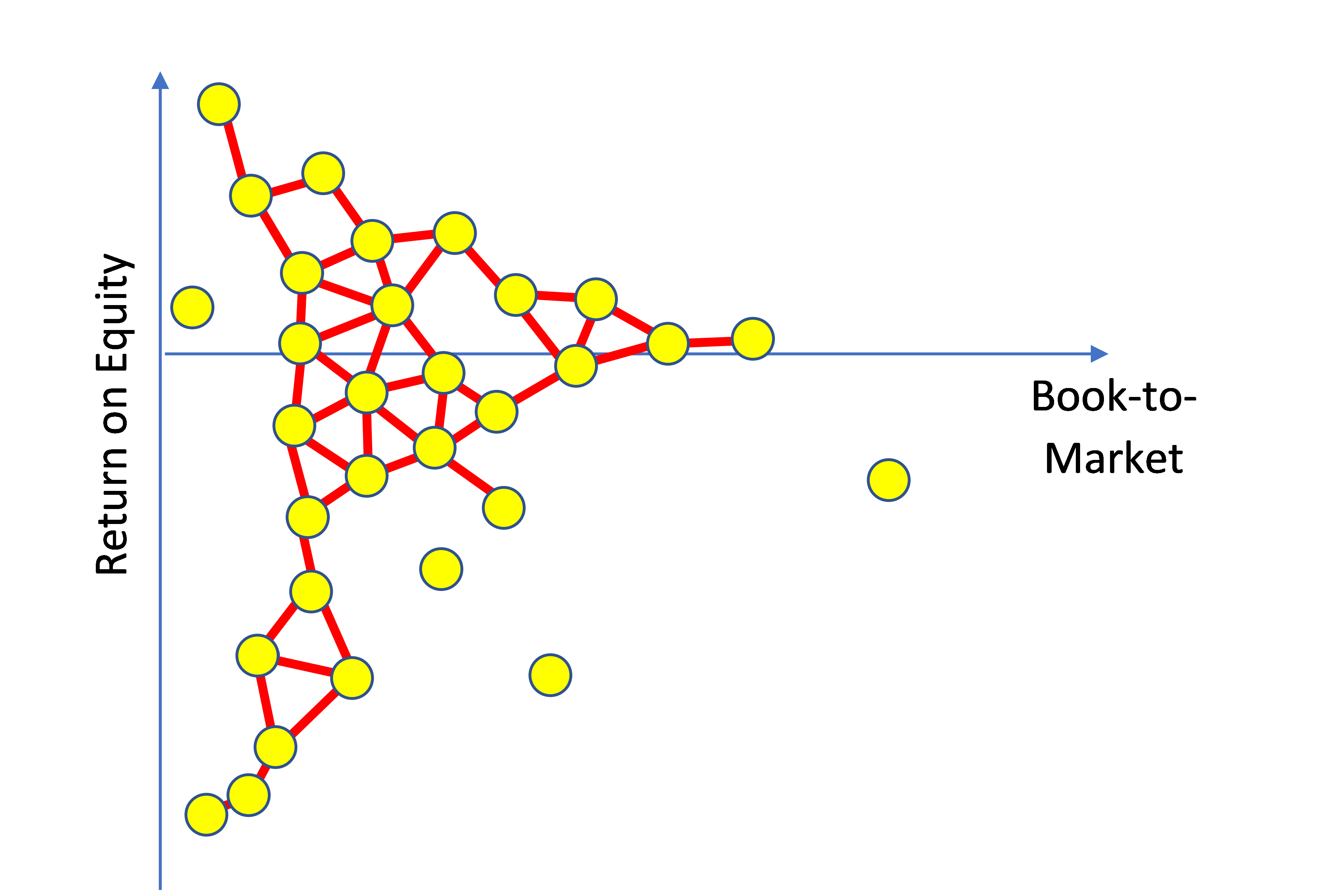}\\
		(c) Mapper graph with data & (d) Abstract mapper graph \\
		\includegraphics[height=4.7cm]{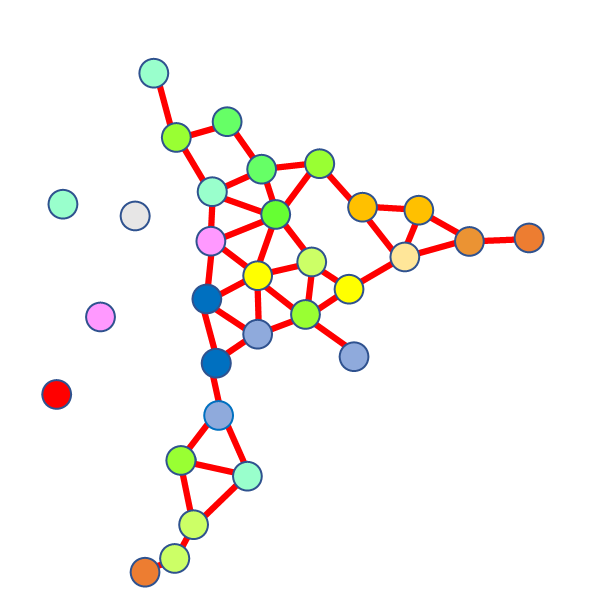}&
		\includegraphics[height=4.7cm]{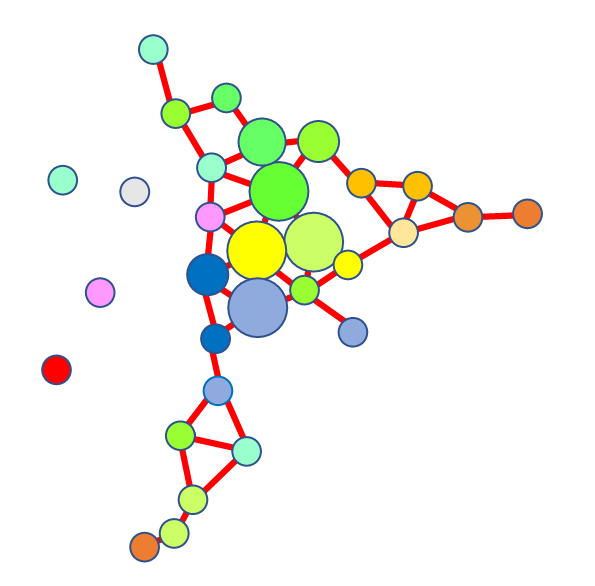}\\
		(e) Colouration of plot & (f) Ball sizes \\
	  \end{tabular}
	\end{center}
 \raggedright
\footnotesize{Notes: Demonstration of construction of TDA Ball Mapper graph\citep{dlotko2019ball} which in addition maintains locations within the space. In the generalised algorithm all plots respect the topography of the space but balls are not arranged according to their axis location in this way. Colouration in panels (e) and (f) reflects expected returns but is deliberately abstracted from existing evidence to show how combinations of axes can often evidence new insights. Ball sizes in Panel (f) represent the number of points within the ball showing where in the space has more data intensity. All plots are illustrative only.} 	
\end{figure}

To construct the TDA Ball Mapper graph we begin with a point cloud and select a point at random therefrom. Around the point a ball is drawn with radius $\epsilon$. Panel (a) of Figure \ref{fig:mapeg2} shows this first ball with its centre coloured yellow. A second point is then selected at random from those which are not covered by the first ball and a ball is drawn around that point. This process continues until all points are covered by at least one ball. To understand the links between these balls we draw a connecting edge wherever there are points in the intersection of the two balls. Panel (b) of Figure \ref{fig:mapeg2} shows a first edge being constructed. Continuing the construction of edges creates the TDA Ball Mapper graph. Panel (c) depicts this. We can then highlight the points that formed the balls and delete the remainder. The result is the representation in panel (d) of Figure \ref{fig:mapeg2}. It is clear that the graph has many of the properties of the original dataset, including the sparse areas at the centre of the cloud and the strings leading down to very negative return on equity values, that going up to high return on equity and the string running close to the zero axis with high book-to-market values. The artificial example can be thought of as being ``Y'' shaped with a big group of observations near the join and a number of outliers.

The number of balls within the plot is very much a function of the radius parameter $\epsilon$. Choosing a higher $\epsilon$ means that fewer balls are needed to cover the whole dataset and hence the plot becomes more sparse. In this way the representation becomes much more abstract, giving a sparser view on the data and losing some of the finer detail. 
Reducing $\epsilon$ means smaller balls and hence a much finer representation. Reinforcing the message from the theoretical exposition, choice of radius is important and the researcher is advised to think carefully about their choice and to check alternatives for robustness. 

Visualisation functionalities can be used to help display more information within the TDA Ball Mapper graph. Colouration is used to plot how a variable of interest varies across the distribution. This may be either the outcome, in this paper that means the stock returns or other characteristics not used in the diagram construction. The TDA Ball Mapper graph may also be coloured like residuals from model fits, predictions or prediction errors. In this way it becomes a tool to explore variation in  model fit across the space. Panel (e) of Figure \ref{fig:mapeg2} shows how the balls may be coloured. In this case the aim is illustration only. By looking at movements through the space it is possible to identify what happens to returns as the firm financial ratios change. In what follows it can be seen that the relationship between returns and the seven financial ratios is non-linear, and that there are a number of patterns within the returns. Interpretation is simply a case of saying what happens in each direction. Colours are set on a spectrum from red through yellow and green to blue and mauve. In panel (e) higher values are seen in the central area with very mixed levels in the outliers; the lack of correspondence is deliberate here because it shows how TDA would be able to identify that the highest returns sit in low book-to-market and nearer zero return on equity. Increasing return on equity but not book-to-market moves towards the top left and falls moving towards lower return on equity. Taking any point in the plot a discussion can be had about the expected effect of moving in any direction. To understand the importance of a ball relative to the sample the size may be set proportional to the number of observations. Panel (f) of Figure \ref{fig:mapeg2} and all subsequent plots in this paper, use size scaling as well as the colouration. Further aiding in the \textit{BallMapper} R implementation used for real data is a numbering of the balls. This numbering is the order the algorithm sets them as ball centres and as such is random; the example plot does not replicate numbers.

At this point the reader may realize that the TDA Ball Mapper has some resemblance to a number of clustering methods. The idea of grouping points that are nearby in the vertices of TDA Ball Mapper is a simple clustering as the points from each node of the TDA Ball Mapper graph, for reasonable choice of $\epsilon$, should belong to the same cluster. Also the points in the TDA Ball Mapper graph that are covered by connected vertices that are nearby in the TDA Ball Mapper graph are likely to belong to the same cluster as well. This is however the end of similarities as the output from the TDA Ball Mapper and the output from any clustering method are very different. In the first case we obtain an abstract graph the structure of which approximates to the considered point cloud. In the other case we obtain a labelling of points as belonging to different clusters. As a consequence of this, it is not possible to present a fair comparison  of the TDA Ball Mapper to the commonly used clustering algorithms. Instead an empirical demonstration of the difference is provided in the discussion.

\section{Bivariate Example: Size and Book-to-Market}
\label{sec:bi}

\begin{figure}
    \begin{center}
        \caption{June 2018 Bivariate Example}
        \label{fig:bi1}
        \begin{tabular}{c c}
            \includegraphics[width=6.7cm]{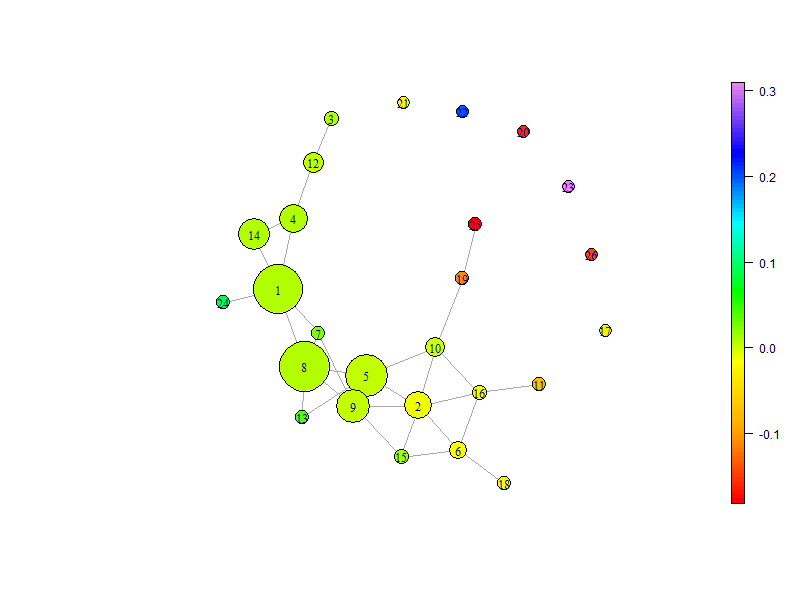} & \includegraphics[width=5.5cm]{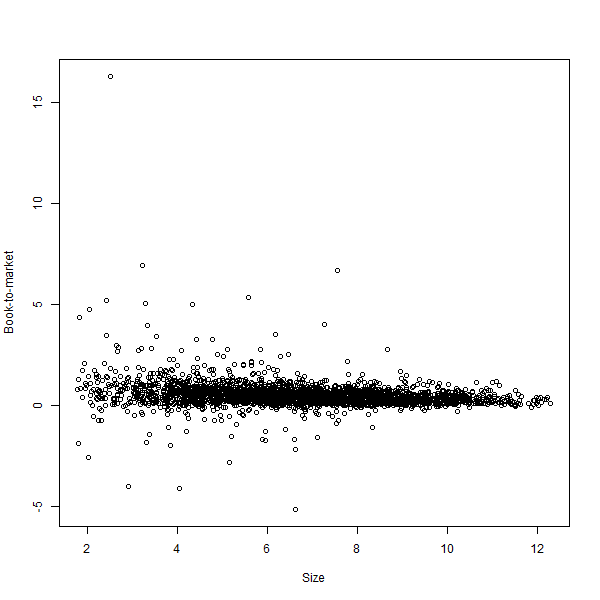}  \\
            (a) Returns Ball Mapper & (b) Scatterplot\\
            \includegraphics[width=6.7cm]{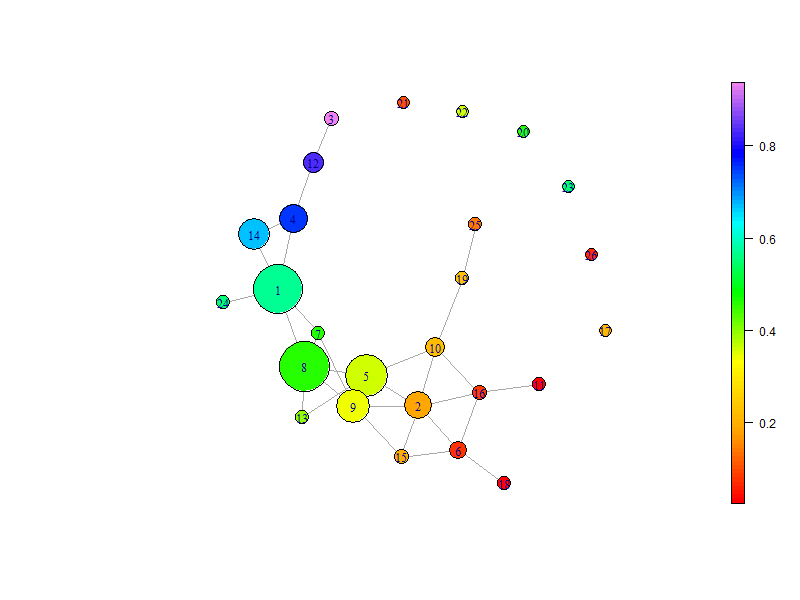}&
			\includegraphics[width=6.7cm]{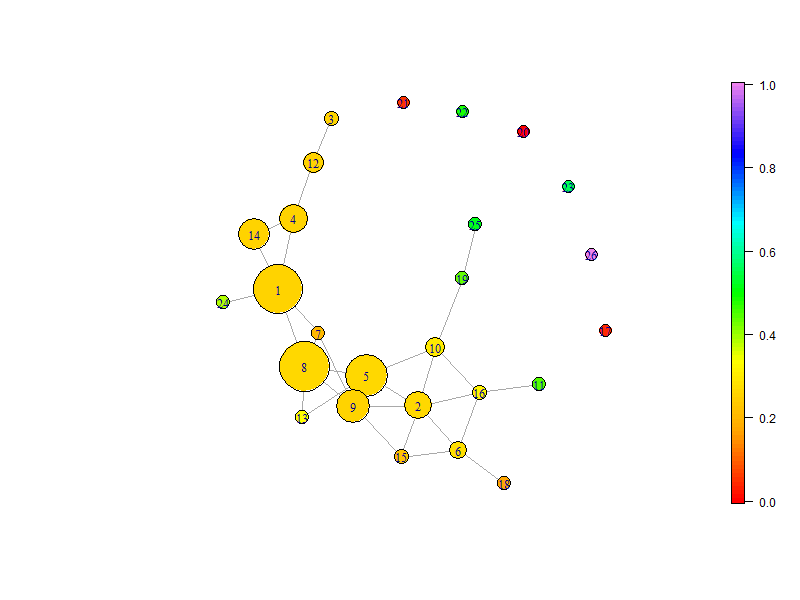}\\
			(c) Size (log) & (d) Book-to-market ratio\\
        \end{tabular}
    \end{center}
\raggedright
\footnotesize{Notes: TDA Ball Mapper plots generated using \textit{BallMapper} \citep{dlotko2019R} with radius $\epsilon=0.1$. Panel (a) coloured by average excess returns for the ball, panel (c) by normalised size and panel (d) by normalised book-to-market ratio. Panel (b) provides scatter plot of data covered. In each case normalisation is to scale $[0,1]$. Colour versions of this plot are available in the online version.}
\end{figure}

\cite{fama1993common} three factor model (FF3) proposes that the CAPM be augmented with sorts on size and book-to-market; a natural bivariate example for the study of firm characteristics is formed. Using \textit{BallMapper} \cite{dlotko2019R} the Ball Mapper algorithm is applied to monthly datasets with two axes, looping over the whole dataset. The monthly datasets are those constructed in the data section and are winzorised for all seven of the axes' variables to facilitate comparability across sections. For brevity only five of the 510 plots are included here; the remainder all share strong similarity. In each the representation of the set is clearly abstract from the scatter plot of the same two variables. Figure \ref{fig:bi1} contrasts the plot for June 2018 with the scatterplot, panel (b), for the same month to show such clearly. In panel (a) colouration is according to the average return and it should be noted that almost all of the values are negative. To this extent ``higher'' simply means less negative. Immediately a crescent shape is seen, with arms which spread at the lower right end. There are a number of outliers around the top right of the plot. Colouration in panel (a) is not consistently evolving from one side to the other; a linear relationship appears inappropriate.

Within \textit{BallMapper} the option to colour the balls according to the axis variables enables ready understanding of how the returns correspond to the respective values of the axes. As each axis is normalised the colour scales on the smaller plots run from 0 to 1; correspondence with actual values can be found in the summary statistics for the plot. Figure \ref{fig:bi1} has a large group of points, centered on a crescent with a number of branches sprouting at the lower right; thinking of the shape thusly allows comparison across plots. Finally six outliers offering very different returns sit to the top right of the plot. Most observations are in the long arm of the shape, starting from balls 5 and 9 on the left of the hexagon. For these plots a small radius $\epsilon=0.10$ is applied as detail about the data is sought. Changing to higher $\epsilon$ values preserves much of the message but presents with less clarity than the version plotted here.

\begin{table}
    \begin{center}
        \caption{Summary Statistics of TDA Ball Mapper Cover: June 2018 ($\epsilon=0.1$)}
         \label{tab:bi1}
        \begin{tabular}{l c c c c c l c c c c }
        \hline
        Ball & Size & BM & Ret (\%) & Obs && Ball & Size & BM & Ret (\%) & Obs\\
        \hline
        1     & 7.724 & 0.390 & 0.742\% & 965&   & 14    & 8.812 & 0.407 & 0.686\% & 491 \\
    2     & 3.770 & 0.593 & -1.156\% & 382  & & 15    & 3.879 & -0.292 & 1.570\% & 37 \\
    3     & 11.588 & 0.237 & 0.672\% & 48    && 16    & 2.724 & 1.677 & -1.923\% & 49 \\
    4     & 9.621 & 0.340 & 1.057\% & 409   & &17    & 4.050 & -4.112 & -3.239\% & 1 \\
    5     & 5.606 & 0.588 & 0.567\% & 785   & &18    & 2.100 & -1.291 & -1.320\% & 5 \\
    6     & 2.636 & 0.692 & -1.514\% & 123   & &19    & 4.205 & 4.140 & -10.994\% & 4 \\
    7     & 6.653 & -0.681 & 2.264\% & 22    & &20    & 6.616 & -5.120 & -15.798\% & 1 \\
    8     & 6.562 & 0.462 & 1.026\% & 1008  && 21    & 2.910 & -3.980 & -2.039\% & 1 \\
    9     & 5.404 & 0.412 & 0.531\% & 534   & &22    & 5.573 & 5.329 & 19.860\% & 1 \\
    10    & 4.111 & 1.239 & -0.062\% & 168   & &23    & 7.549 & 6.659 & 30.674\% & 1 \\
    11    & 2.175 & 4.448 & -7.415\% & 4     & &24    & 7.522 & 3.078 & 8.772\% & 2 \\
    12    & 10.538 & 0.337 & 0.364\% & 197   & &25    & 3.251 & 5.990 & -17.825\% & 2 \\
    13    & 5.905 & 2.103 & 3.824\% & 17    & &26    & 2.517 & 16.268 & -14.034\% & 1 \\

    \hline
        \end{tabular}
        
    \end{center}
\raggedright
\footnotesize{Notes: Summary statistics calculated for Size (log of market value) and BM (book-to-market ratio). Returns expressed in percentage form. Obs records number of observations in each ball. Recall that points may appear in multiple balls. }
\end{table}

Table \ref{tab:bi1} gives summary statistics for the 26 Balls that feature in Figure \ref{fig:bi1}. To the top left of the crescent lie the largest firms; moving through balls 4 to 12 to 3 sees size increase and modest reductions in book-to-market. In this area there is a fall in the return percentage from 1.057\% in ball 4 to 0.364\% in ball 12 then back up to 0.672\% in ball 3. By contrast at the other end of the conjoined shape the rays relate to much smaller clusters of firms. Just 6 stocks exist on the ray 10 to 19 to 25, a ray with small size and mid level book-to-market. The shorter arm to ball 11 from 16 sees a movement towards very small sized firms and, once again, a slight fall in book-to-market. Finally the short arm from ball 6 to ball 19 is associated with very small firms once more, and this time a lower book-to-market. Using Table \ref{tab:bi1} the returns on these paths can be shown to be falling significantly on the first two, ball 25 has an average return of -17.83\% and ball 11 an average excess return of -7.415\%. By contrast ball 18 has much higher returns, the average value being just -1.32\%. In a month where there were many losses both ends of the shape show a big difference in a small part of the parameter space.

\begin{figure}
	\begin{center}
	    \caption{Evolution of Bivariate Plots}
	    \label{fig:bi2}
		\begin{tabular}{c c c c}
				\multicolumn{2}{c}{JUNE 1978} & \multicolumn{2}{c}{JUNE 1988}\\
			\multicolumn{2}{c}{\includegraphics[width=6.5cm]{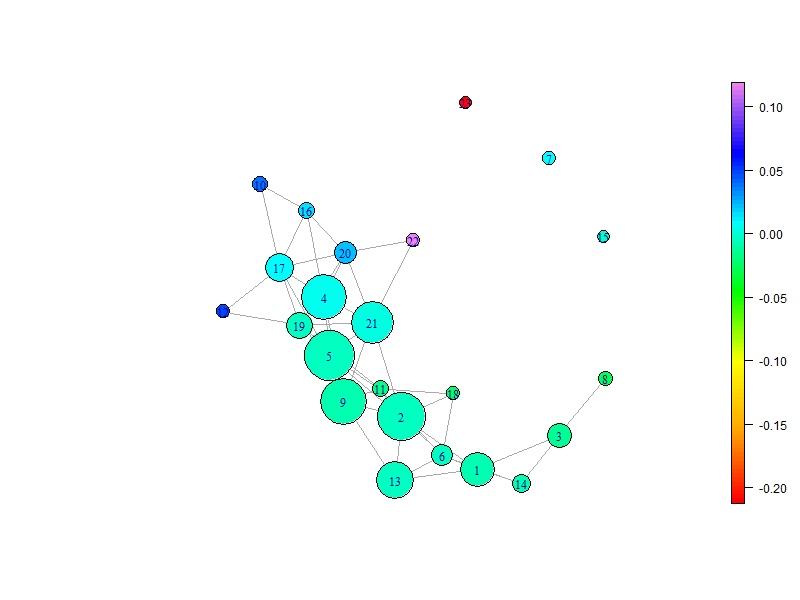}}&
			\multicolumn{2}{c}{\includegraphics[width=6.5cm]{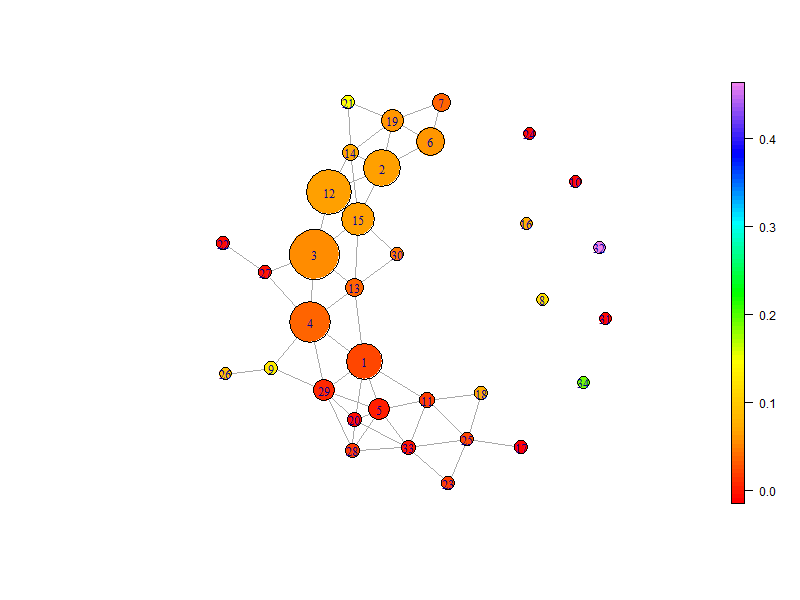}}\\
			\multicolumn{2}{c}{ (a) Returns}&\multicolumn{2}{c}{ (b) Returns}\\
			\includegraphics[width=3.2cm]{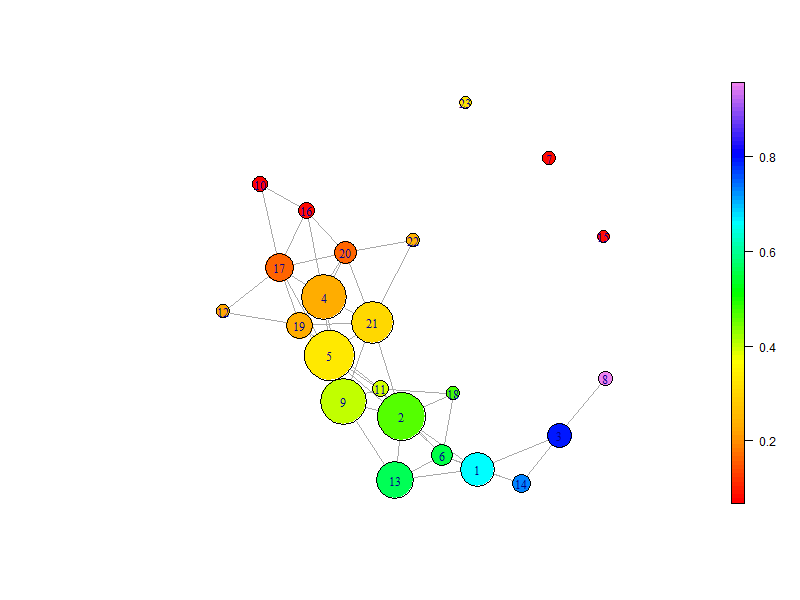}&
			\includegraphics[width=3.2cm]{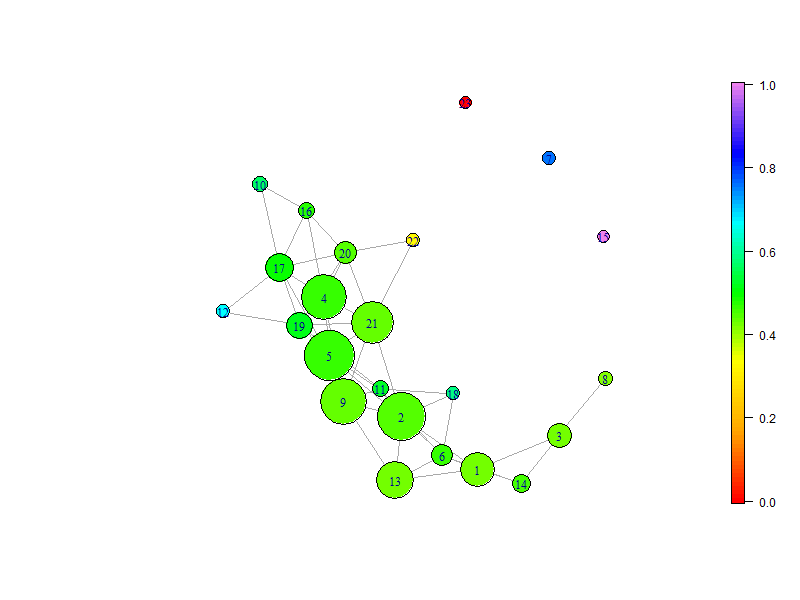}&
			\includegraphics[width=3.2cm]{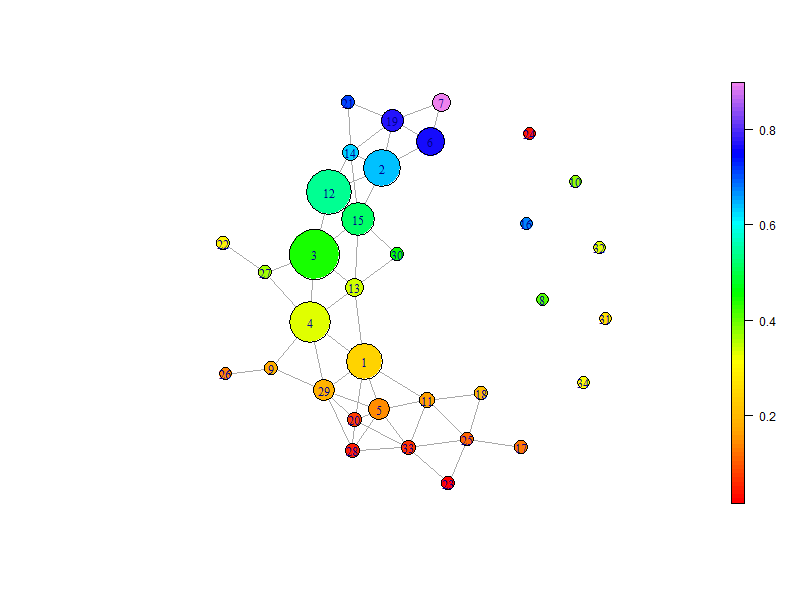}&
			\includegraphics[width=3.2cm]{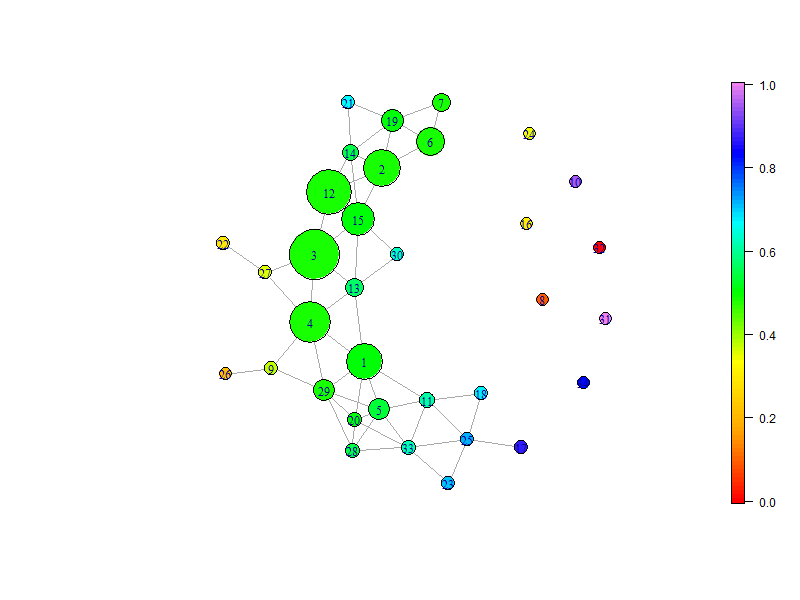}\\
			(c) Size & (d) Book-to-market & (e) Size & (f) Book-to-market \\
			\vspace{10pt}
			\\
			\multicolumn{2}{c}{JUNE 1998} & \multicolumn{2}{c}{JUNE 2008}\\
			\multicolumn{2}{c}{\includegraphics[width=6.5cm]{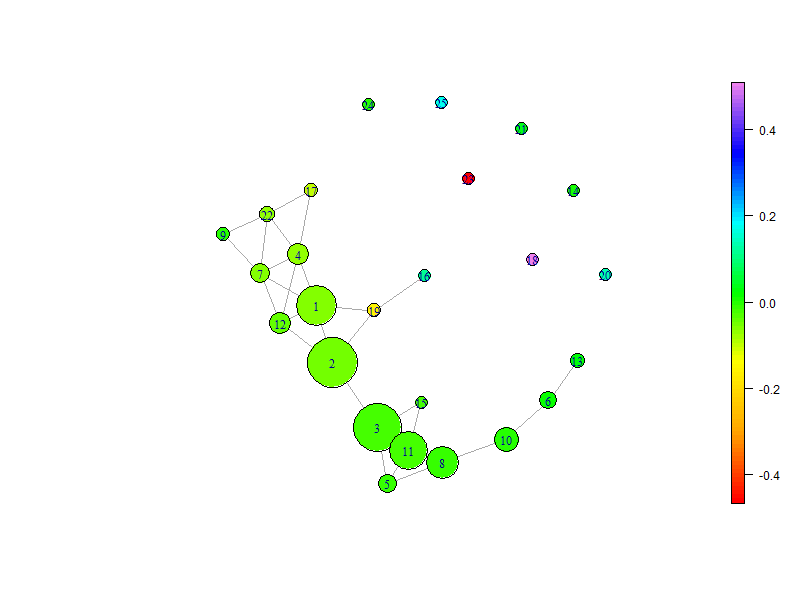}}&
			\multicolumn{2}{c}{\includegraphics[width=6.5cm]{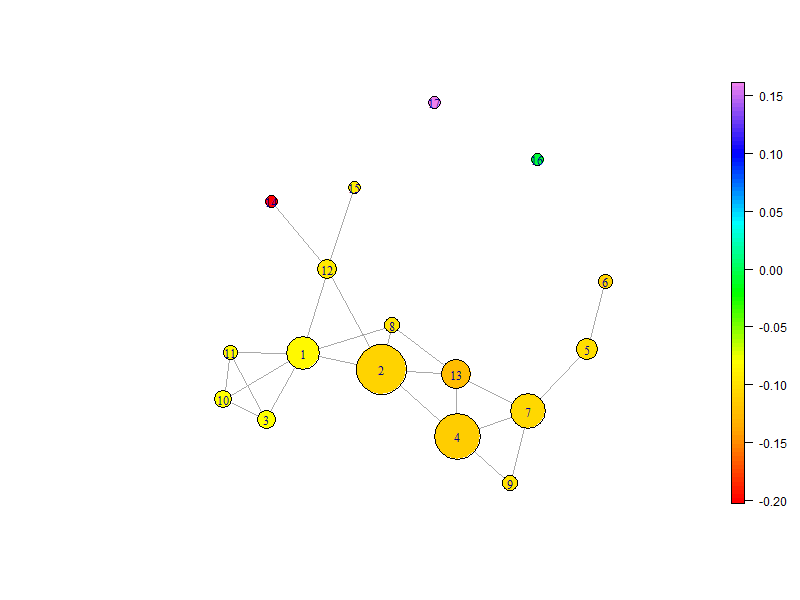}}\\
			\multicolumn{2}{c}{(g) Returns}&\multicolumn{2}{c}{(h) Returns}\\
			\includegraphics[width=3.2cm]{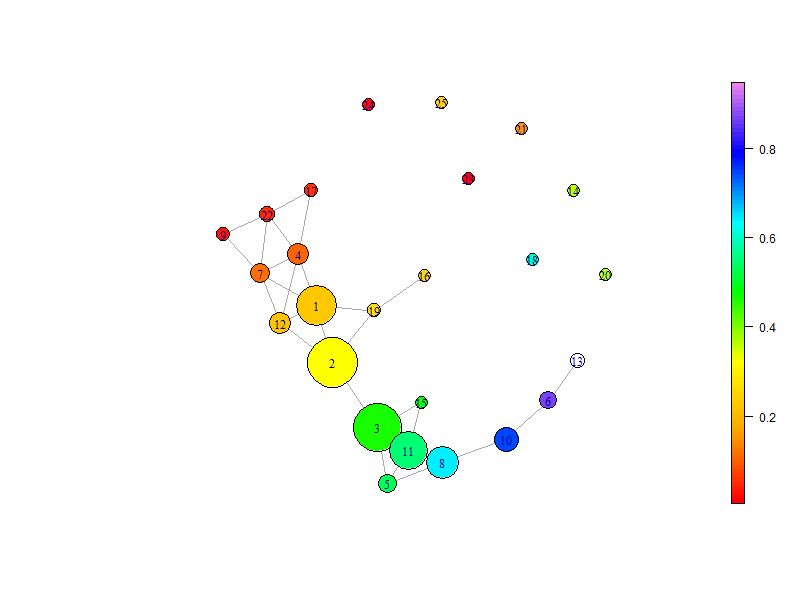}&
			\includegraphics[width=3.2cm]{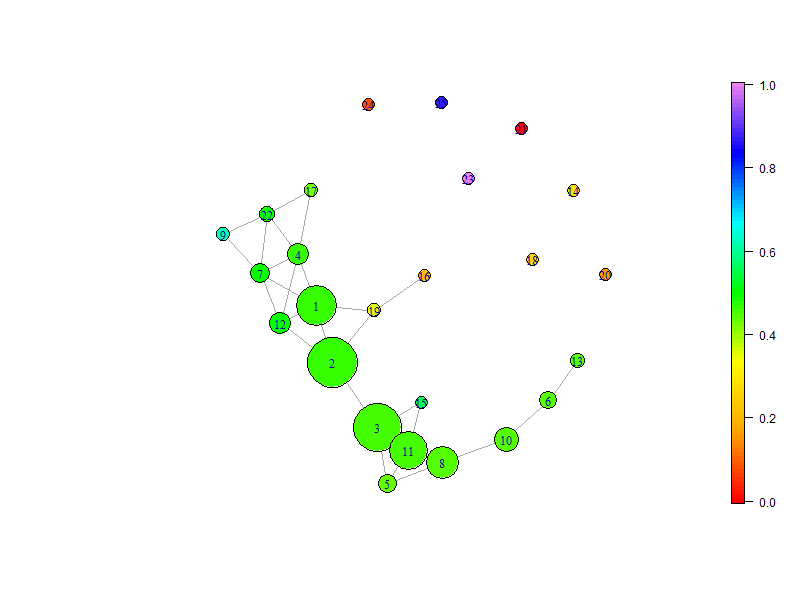}&
			\includegraphics[width=3.2cm]{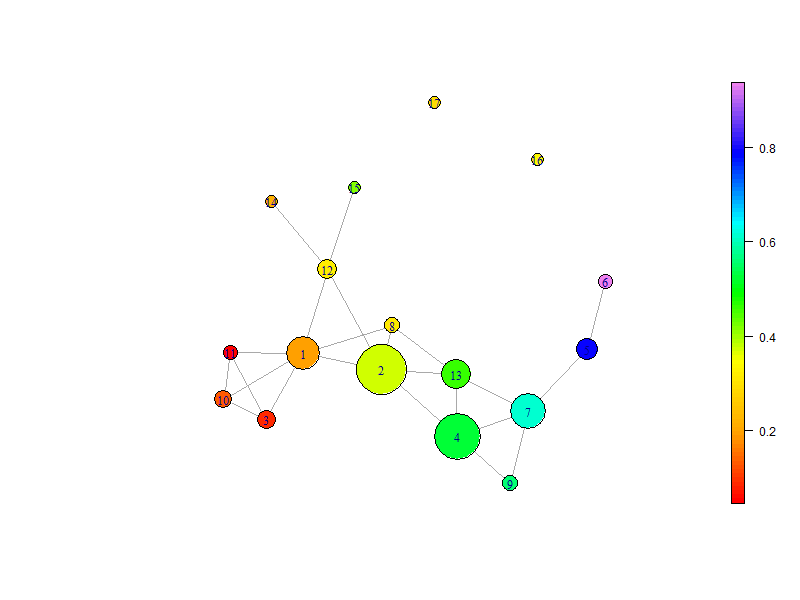}&
			\includegraphics[width=3.2cm]{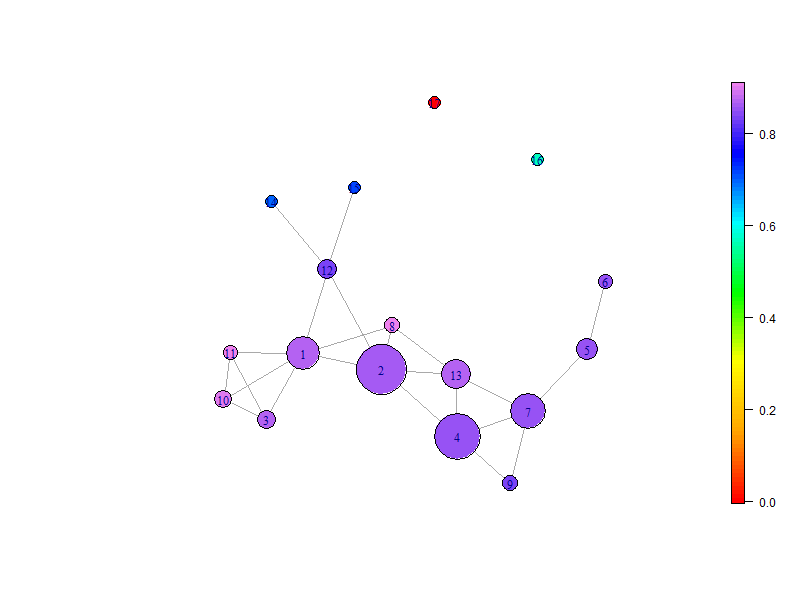}\\
			(i) Size & (j) Book-to-market & (k) Size & (l) Book-to-market \\
			
		\end{tabular}
	\end{center}
\raggedright
\footnotesize{Notes: TDA Ball Mapper plots generated using \textit{BallMapper} \citep{dlotko2019R} with radius $\epsilon=0.1$. Capitalised headers inform to which year the plots below relate. Panels (a), (b), (g) and (h) are coloured by average excess returns for the ball. Panels (c), (e), (i) and (k) are coloured by normalised size. Panels (d), (f), (j) and (l) are coloured by normalised book-to-market ratio. In each case normalisation of the two axis variables is to scale [0,1]. Colour versions of this plot are available in the online version.}
\end{figure}

To understand the consistency of this relationship across time, Figure \ref{fig:bi2} gives the corresponding bivariate plots for June 1978, 1988, 1998 and 2008. These were very different years in terms of the state of the market at the time; 2008 being the deepest part of the Global Financial Crisis (GFC). June 1978 was a period with a climbing stock market, checked only in the final few days. June 1988 came after the 1987 ``Black Wednesday'' crash and was again a period of growth in the market. June 1998 market the end of a bear market that had seen the S\&P 500 index climb 21\% in just over 12 months\footnote{All observations on years from market index behaviours with data from CRSP.}. Across the panels of Figure \ref{fig:bi2} the broad crescent shape features strongly, with the outlying set being clearly visible. In June 1978 there are fewer outliers, representing a lower dispersion of firms within the sample. The spread at the end of the crescent is still present but the arms are much less pronounced than in 2018. A reminder that these plots are abstract comes in the fact that the split is at the left end of the crescent rather than the right. June 1988 still has a greater spread resulting in a bigger net throughout the crescent. There are seven outliers reinforcing the spread message. Notable here is that the lowest, red, balls correspond to returns around 0\%. June 1998 sees a lot of negative returns, with only one of the outlier balls coloured as having a positive average. June 1998 also returns to the spread being at the left of the crescent. In 2008 the majority of the outliers connect to the main shape and the arms are much more pronounced. Once again the arms come at the left end of the plot. 

Links to the underlying characteristics can be found in the smaller plots below. In each case size varies monotonically across the main connected shape, the smallest firms being where the shape becomes the most spread. Book-to-market ratios vary less because the outliers contain the extreme values. Using the plots, the same relationship of arms with low book-to-market and small size, mid-range book-to-market and small size, and mid-range book-to-market and mid-range size are seen in most cases. Only June 1998 does not show such a pattern. Because the more extreme values of book to market vary over the time the actual colour used for the plot also changes; in every case the range of colouration within the main shape is small however. A full summary table is not reported here for brevity but is available upon request. 

A particular benefit of the TDA Ball Mapper approach is that all possible combinations of characteristics may be included. Through the plot it is apparent that the union of large size and high book-to-market values comes in the thin arm of the shape and in the outliers, it does not however occur at the end of the shape. In June 2018 this union was associated with the mid-range of returns, losses of around 0.05\% (Figure \ref{fig:bi1}). Likewise larger firms with higher book-to-market ratios were returning around -0.05\% in June 2008 (Figure \ref{fig:bi2} panels (g), (j) and (k)). Such results are readily verified in a cross tabulation of sorts of the stocks on the two characteristics\footnote{This is the basis of the two way sorts used as test assets for model comparison \citep{fama2018choosing}.}. Such tabulation of two variables is straightforward, but again this quickly becomes complex as the number of dimensions is increased.

Exposition of the bivariate case provides a neat introduction to the ability of TDA Ball Mapper to represent data. From the outset it is clear that relationships between the key characteristic variables from the FF3 models and returns are not linear. Through the plots clarity on the role of combinations of the two variables was also found. These joint-effects are easy to bring into cross-tabulations and regressions as an interaction term, but such would quickly become problematic as the set of explanatory factors enlarges. All that has been presented in two dimensions is intuitive, but it is to the expansion of the number of dimensions that TDA Ball Mapper really speaks. 

\section{Common Anomalies}
\label{sec:sort}

In the data section seven common anomalies were discussed in their financial ratio form. These anomalies form a seven dimensional point cloud to which the TDA Ball Mapper algorithm is applied. This is again performed with implementation in R using \cite{dlotko2019R}. As in the bivariate case data is taken from the winzorised monthly subsamples. Unlike the bivariate case it is not easy to visualise the data using simple scatter plots rendering the TDA Ball Mapper graphs completely abstract representations. Thus the value of being able to colour by axes is clear; as will be demonstrated, from thence stems the discussion of individual characteristics. 

\subsection{TDA Ball Mapper Plots}

\begin{figure}
	\begin{center}
	    \caption{Common Anomalies TDA Ball Mapper Plots: June 2018}
	    \label{fig:multi1}
		\begin{tabular}{c c c}
			\multicolumn{3}{c}{\includegraphics[width=12cm,height=8.5cm]{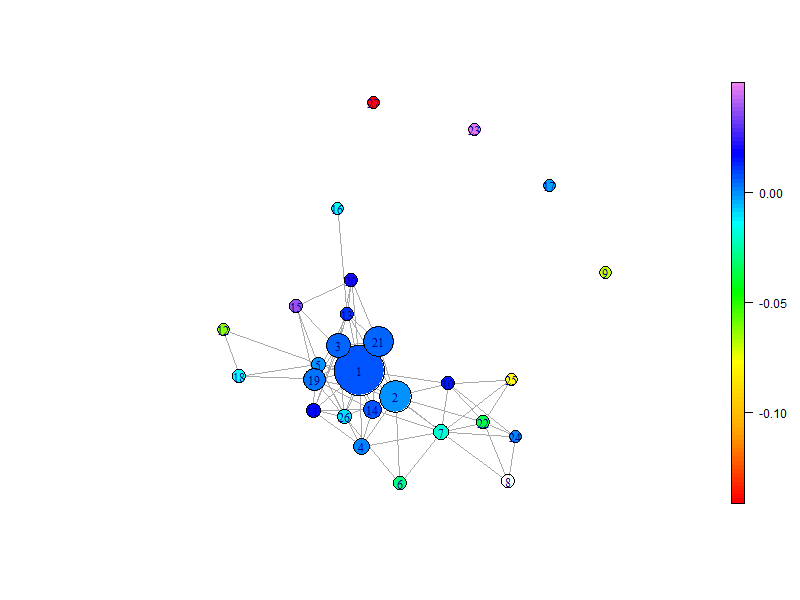}}\\
			\multicolumn{3}{c}{(a) Returns}\\
			\includegraphics[width=4.2cm]{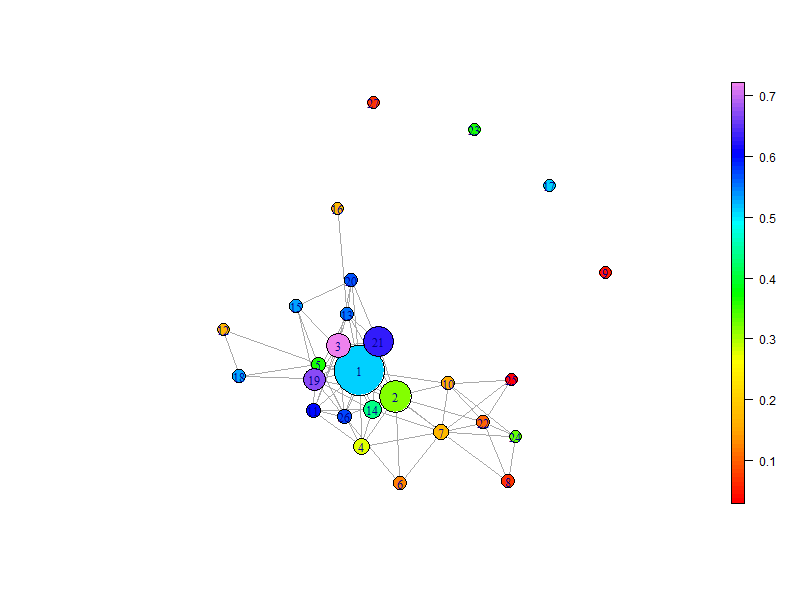}&
			\includegraphics[width=4.2cm]{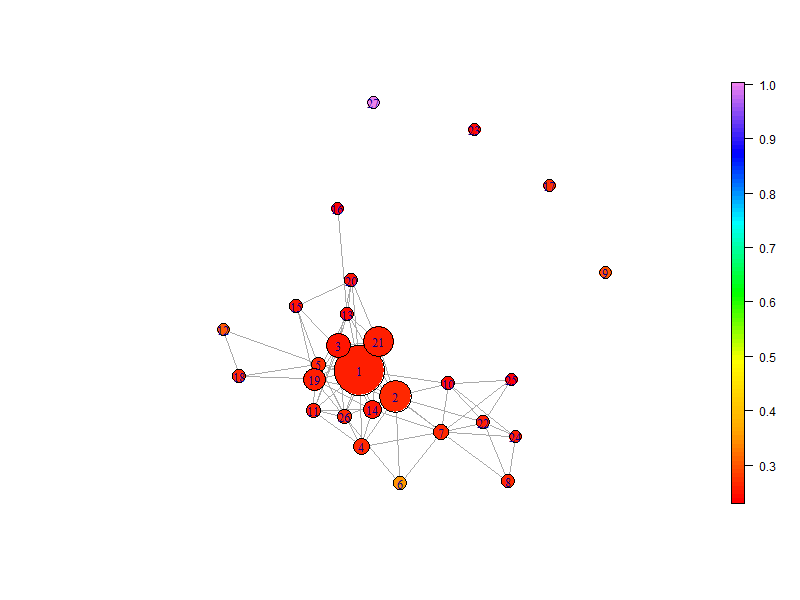}&
			\includegraphics[width=4.2cm]{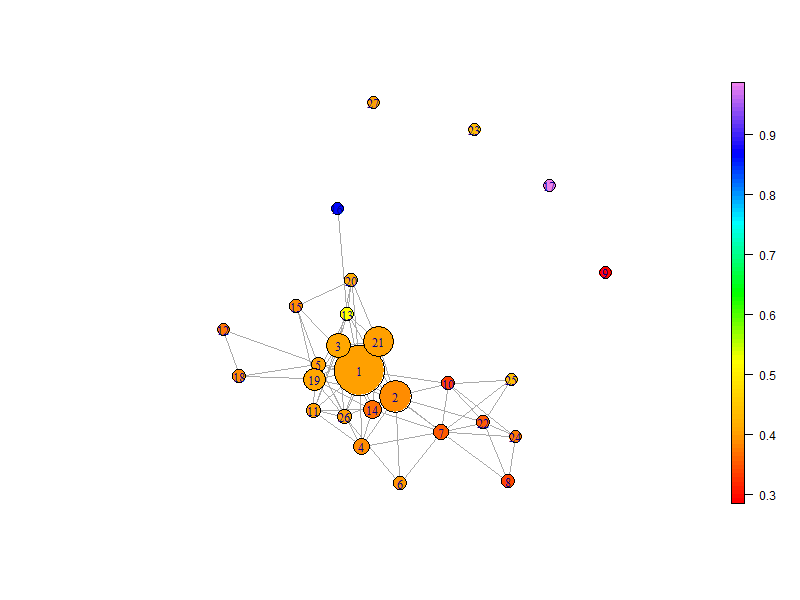}\\
			(b) Size & (c) Book-to-market & (d) Profitability \\
			\includegraphics[width=4.2cm]{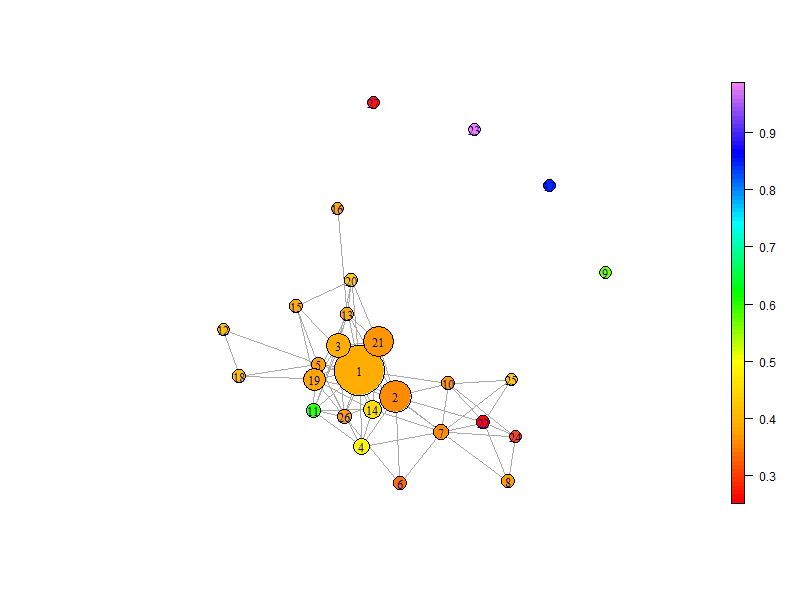}
			&
		    \includegraphics[width=4.2cm]{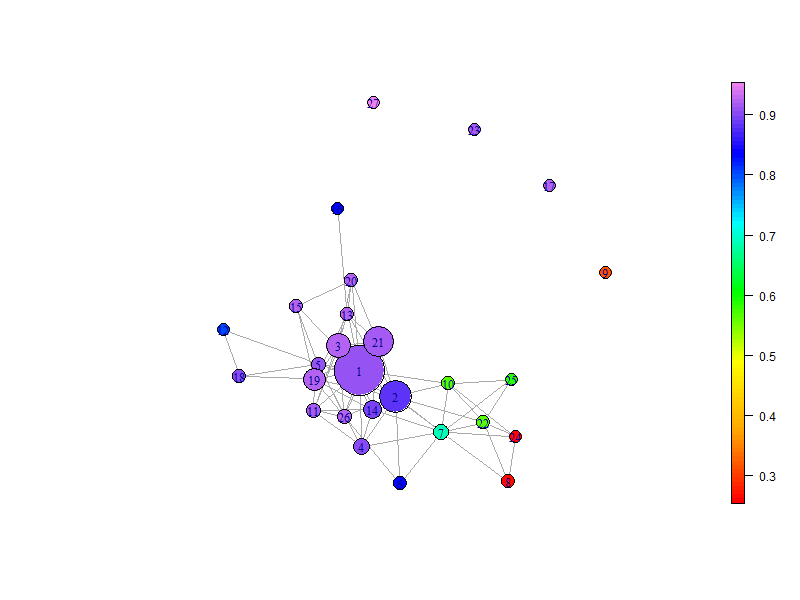}&
			\includegraphics[width=4.2cm]{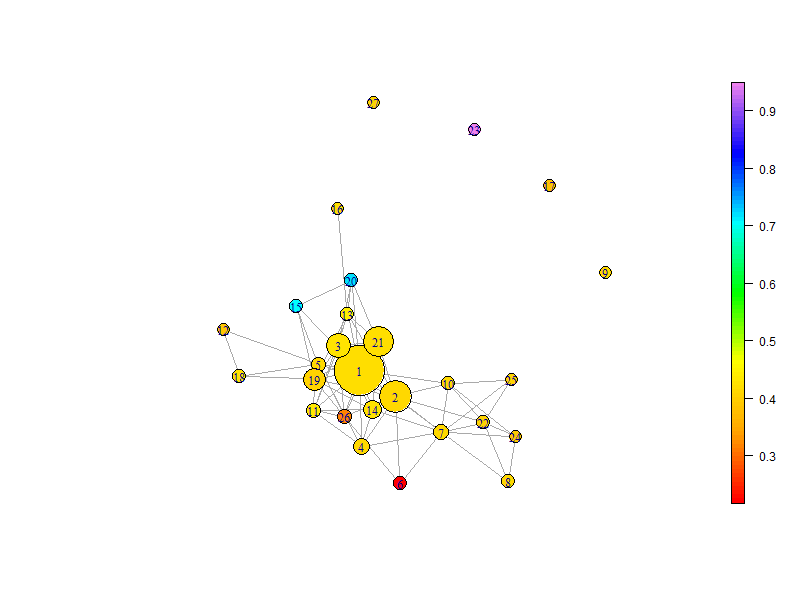}
			\\
			(e) Investment & (f) Earnings-to-price & (g) Cash-to-price \\
			
			\includegraphics[width=4.2cm]{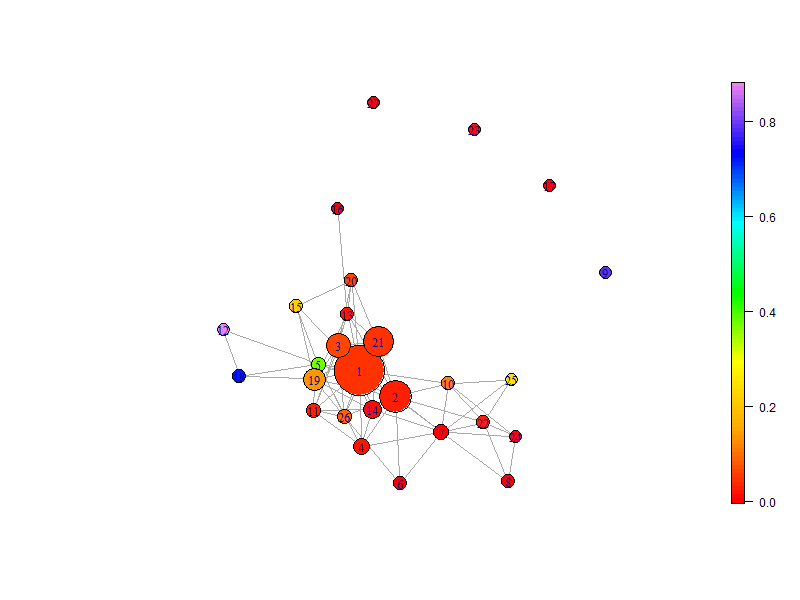}&
			\includegraphics[width=4.2cm]{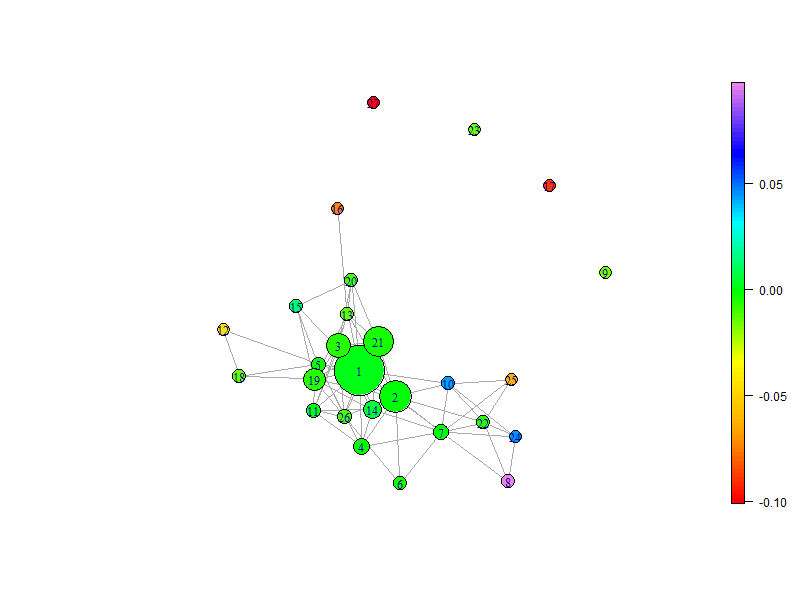}&
			\includegraphics[width=4.2cm]{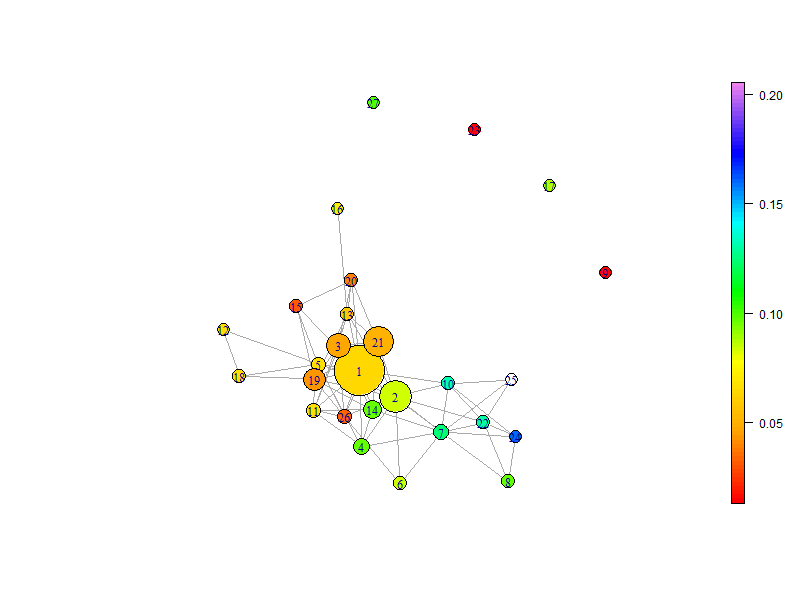}\\
			(h) Dividend yield & (i) Residuals & (j) Absolute Residuals \\
		\end{tabular}
	\end{center}
\raggedright
\footnotesize{Notes: TDA Ball Mapper plots generated using \textit{BallMapper} \citep{dlotko2019R} with radius $\epsilon=0.40$. Panel (a) coloured by average excess returns for the ball, panels (b) to (h) coloured by normalised values of the variables named below. In each case normalisation is to scale [0,1]. Residuals, panels (i) and (j), from OLS regressions of returns on the seven axis characteristics using June 2018 data. Colour versions available online.}
\end{figure}

A particular strength of TDA Ball Mapper is the facilitation of discussion of combinations of characteristics. Again the axes plots help greatly with this. Figure \ref{fig:multi1} demonstrates. Panel (a) shows that the introduction of seven axes has not led to the creation of a large set of diverging lines. To the left of the mass lie two balls which are well connected but which have quite different returns. To the upper side returns generally increase, the colouration darkens, but a single ball is seen with negative returns. To the lower right there is an extended net which includes predominantly high returns. However, there are also two balls, connected into the set, which have strong negative returns despite sharing so many characteristics. Outside of the connected set are four outlier balls, some with high returns and others with very low returns. Recalling that a connection between balls only forms when they are sufficiently close in the space and have points in the intersection, the connection of very low returning balls to high returners is of deep interest.

\begin{table}
    \begin{center}
        \caption{Summary Statistics for Seven Common Anomalies Data: June 2018 ($\epsilon=0.40$)}
        \label{tab:multi1}
        \begin{small}
        \begin{tabular}{l c c c c c c c c c}
             \hline
             Ball & Size & BM & ROE & Invest & EP & Cash & DY & Returns & Obs \\
             \hline
                 1     & 7.171 & 0.463 & 0.006 & 0.029 & 0.000 & 2.245 & 1.102\% & 0.704\% & 2438 \\
    2     & 5.114 & 0.590 & -0.196 & 0.001 & -0.099 & -3.734 & 0.667\% & -0.126\% & 1246 \\
    3     & 9.338 & 0.357 & 0.150 & 0.030 & 0.036 & 7.365 & 1.415\% & 0.520\% & 727 \\
    4     & 4.700 & 0.687 & -0.158 & 0.135 & -0.032 & -0.239 & 0.422\% & 0.120\% & 203 \\
    5     & 5.630 & 0.677 & -0.007 & 0.012 & -0.013 & -2.260 & 8.256\% & -0.206\% & 96 \\
    6     & 3.068 & 2.460 & -0.103 & -0.024 & -0.274 & -117.093 & 0.017\% & -2.982\% & 14 \\
    7     & 3.627 & 0.657 & -0.898 & -0.001 & -0.765 & -1.992 & 0.154\% & -2.044\% & 189 \\
    8     & 2.518 & 0.673 & -1.087 & 0.011 & -2.194 & -2.035 & 0.000\% & 4.886\% & 9 \\
    9     & 2.329 & 1.231 & -2.031 & 0.207 & -2.035 & -0.837 & 17.606\% & -7.022\% & 1 \\
    10    & 3.445 & 0.137 & -1.211 & 0.003 & -1.188 & -2.643 & 2.640\% & 1.413\% & 22 \\
    11    & 8.053 & 0.376 & 0.082 & 0.238 & 0.013 & 8.621 & 0.681\% & 1.490\% & 104 \\
    12    & 3.606 & 1.487 & -0.460 & 0.048 & -0.372 & -16.11 & 19.61\% & -6.459\% & 6 \\
    13    & 7.545 & 0.114 & 2.170 & 0.030 & 0.040 & 17.14 & 0.439\% & 1.024\% & 40 \\
    14    & 6.311 & 0.365 & -0.587 & 0.096 & -0.057 & 6.090 & 0.240\% & 0.723\% & 380 \\
    15    & 7.360 & 0.223 & -0.155 & 0.033 & 0.044 & 172.3 & 4.762\% & 3.629\% & 12 \\
    16    & 3.619 & -0.123 & 8.418 & 0.025 & -0.265 & 3.620 & 0.000\% & -1.724\% & 3 \\
    17    & 7.152 & 0.702 & 10.561 & 0.493 & 0.041 & -21.50 & 0.000\% & -0.360\% & 1 \\
    18    & 7.390 & 0.665 & -0.174 & 0.052 & -0.021 & 4.377 & 16.08\% & -1.450\% & 10 \\
    19    & 8.824 & 0.443 & 0.143 & 0.020 & 0.042 & -4.649 & 2.973\% & 0.236\% & 586 \\
    20    & 7.686 & 0.223 & 0.260 & 0.064 & 0.033 & 179.77 & 1.430\% & 1.949\% & 39 \\
    21    & 8.372 & 0.393 & 0.087 & 0.004 & 0.022 & 3.634 & 1.083\% & 0.406\% & 1113 \\
    22    & 2.906 & 0.416 & -0.885 & -0.110 & -1.185 & -5.356 & 0.770\% & -3.739\% & 26 \\
    23    & 5.646 & 0.114 & 0.622 & 0.640 & 0.011 & 316.0 & 0.367\% & 4.535\% & 1 \\
    24    & 5.249 & 0.518 & -0.496 & -0.055 & -2.239 & -13.17 & 0.000\% & 0.047\% & 6 \\
    25    & 2.127 & -0.087 & 0.735 & 0.063 & -1.108 & -7.516 & 5.324\% & -7.734\% & 7 \\
    26    & 7.778 & 0.638 & 0.091 & 0.017 & 0.044 & -57.17 & 2.114\% & -1.059\% & 134 \\
    27    & 2.517 & 16.268 & 0.009 & -0.092 & 0.147 & -5.274 & 0.000\% & -14.03\% & 1 \\
    \hline
        \end{tabular}
        \end{small}
    \end{center}
\raggedright
\footnotesize{Notes: Summary statistics are obtained using R and the coverage functionality in \textit{BallMapper} \citep{dlotko2019R}. Size is the log of market valuation, BM the book-to-market ratio, ROE the return-on-equity, Invest is investment to asset ratio, EP is the earnings-to-price ratio, Cash the cashflow-to-price ratio, and DY the dividend yield. Returns are the excess returns for the ball. Dividend yield and returns are expressed as percentages. Obs is the number of observations within each ball.}
\end{table}

Colouration of the axes in the seven panels below the main plot is helpful to interpreting the output in panel (a). These plots are coloured according to their values on each of the axes, with the values normalised on the interval $[0,1]$. Discussion of the colouration is aided by reference to Table \ref{tab:multi1}, which includes summary statistics for all 27 of the balls in the plots. Panel (b) shows size to be largest in the area immediately above the common mass. Comparison with panel (a) shows that many of these balls also have higher returns, seemingly going against the assumption that larger firms have smaller returns. Small firms in the lower right net generate some of the greatest returns, but also some of the lowest outwith the outliers. From the discussion of size the importance of joint-effects already emerges.

Book-to-market is visualised in panel (c), with many observations fitting in a narrow range. Whilst differentiating precisely between balls becomes hard in such cases it is clear that the higher values are in the extremes of the plot, ball 6 for example. In the returns plot ball 6 is a lower returning ball. Likewise the outlier with the highest book-to-market is the lowest returner. Both results suggest that higher book-to-market would mean lower returns, going against the established order. In this example there is motivation to use TDA Ball Mapper to look at what other characteristics these firms have. Ball 6 is shown in Table \ref{tab:multi1} to have 14 constituents so is worth a bit more thought. The very low return outlier, ball 25, has 7 members so should not be dismissed. In the bivariate case Figure \ref{fig:bi1} showed mixed results also, but broadly higher book-to-market was associated with balls that had lower returns. 

\cite{fama2015five} (FF5) add profitability and investment as anomalies not covered by their original FF3 model. Plot (d) shows the lower right net to be the area with the lowest profitability; these are already noted to be smaller firms. There is some evidence of high profit in the firms above the main cluster also; this arm had some large firms. Returns in the former group are lower, whilst in this latter set returns are much higher. Herein some non-linearity is evidenced. Investment is the final of the FF5 sorts, with panel (e) showing that lower levels of investment are found most consistently in the area just below the big central balls. Here returns are more variable, but are typically lower than in the central balls. A similar story can be told about the net area to the lower right of the plot. This is again inconsistent with the conservative (low investment) minus aggressive (high investment) factor used in FF5.

Three additional firm characteristics are considered within this paper, and these start to combine to produce an explanation for the observed patterns within the data. Earnings-to-price in panel (f) is also lower in the spread net area of the plot where profitability was also lower; this is an inevitable feature for variables with high correlation. However, other lower earnings-to-price groups do not align with balls where profitability had differed from the main cluster. Cash-to-price, panel (g), is also higher in the upper areas of the plot and there are some lower values below the main mass. In the net area there is no notable differentiation in the cash-to-price relative to the central area. Recalling, however, that the TDA Mapper plot is an abstract representation it may be inferred that despite the distance on the diagram many of these extreme values may actually be close to connecting with each other. Finally the dividend yield, panel (h), is at its highest in the left area. Recalling that most firms did not pay dividends, the median was zero; this lack of variation is unsurprising. That such a wide variety of dividend yields are seen in the same area follows from the fact that TDA Ball Mapper would segregate balls that were otherwise equal in other dimensions as they would be more than $\epsilon$ apart in dividend yield. Once again this creates a situation where the returns are dependent on the combination of dividend yield and one (or more) of the other variables. 

Driving the discussion from the individual axes is useful to say something about how each variable affects the outcomes. Many have shown that in fact sometimes their highest values are associated with high returns, but other times they are not; the discussion around Figure \ref{fig:multi1} leads to a rejection of the notion of monotonic relationships. Combining all of the above it is possible to then form an opinion about what brings high returns in June 2018. The message in its simplest form is that it is large firms with average book-to-market ratios, strong profitability, with limited investment, but who have strong earnings-to-price, higher cash-flows. The high returning stocks will do so, but with a comparatively low dividend yield. This is based on the area of the plot immediately above the main cluster.

From the plots the main message is just how compact the distributions of many of the axis variables are. Trade-offs must therefore be made between reducing $\epsilon$ to obtain more balls away from the central core, and doing further winzorisation so that the central core is spread more by the normalisation process. As an unreported experiment both have been performed but the results remain qualitatively similar. The takeaways of non-linearity and the importance of combinations of factors still shine strong.

\subsection{Post Estimation}

The \textit{BallMapper} package \citep{dlotko2019R} contains a number of useful functions to facilitate more detailed analysis of the obtained graphs. Many questions are raised when looking at panel (a) of Figure \ref{fig:multi1}. Investors may be interested in why the firms covered by the ball 26 delivers, on average, returns so much lower than the neighbouring balls. Likewise there are low returns found at the top and bottom of the plot, but most axis variables appeared the same. Why are these balls not connected?. These questions require the ball comparison functionality of the \textit{BallMapper} \citep{dlotko2019R} package. 

\begin{table}
    \begin{center}
    \begin{tiny}
        \caption{Ball Comparisons for June 2018}
        \label{tab:comp1}
        
        \begin{tabular}{l l c c c c c c c c c c }\hline
            \multicolumn{2}{r}{Compare:} & \multicolumn{2}{c}{26} & \multicolumn{2}{c}{26} & \multicolumn{2}{c}{25} & \multicolumn{2}{c}{25} & \multicolumn{2}{c}{6} \\
             \multicolumn{2}{r}{With:} & \multicolumn{2}{c}{11} & \multicolumn{2}{c}{1,4,11,14} & \multicolumn{2}{c}{12} & \multicolumn{2}{c}{10} & \multicolumn{2}{c}{16}  \\
             & & Diff. & Dist & Diff. & Dist & Diff. & Dist & Diff. & Dist & Diff. & Dist\\
             \hline
        \multicolumn{2}{l}{Size} & -0.274 & 0.131 & 0.695 & 0.333 & -1.479 & 0.707 & -1.318 & -0.630 & -0.551 & -0.264 \\
        \multicolumn{2}{l}{Book-to-market} & 0.262 & 0.413 & 0.168 & 0.265 & -1.574 & -2.475 & -0.223 & -0.351 & 2.583 & 4.063 \\
        \multicolumn{2}{l}{Profitability} & 0.008 & 0.010 & 0.107 & 0.132 & 1.195 & 1.469 & 1.946 & 2.392 & -8.521 & -10.48 \\
        \multicolumn{2}{l}{Investment} & -0.221 & -2.760 & -0.015 & -0.192 & 0.015 & 0.190 & 0.060 & 0.753 & -0.049 & -0.611 \\
        \multicolumn{2}{l}{Earnings-to-price} & 0.030 & 0.109 & 0.047 & 0.169 & -0.736 & -2.662 & 0.080 & 0.290 & -0.009 & -0.033 \\
        \multicolumn{2}{l}{Cashflow-to-price} & -65.79 & -1.879 & -59.34 & -1.694 & 8.591 & 0.245 & -4.873 & -0.139 & -12.07 & -3.447\\
        \multicolumn{2}{l}{Dividend Yield} & 0.014 & 0.661 & 0.010 & 0.474 & -0.143 & -6.558 & 0.027 & 1.238 & 0.000 & 0.008\\
        \hline
        \end{tabular}
    \end{tiny}
    \end{center}
\raggedright
\footnotesize{Notes: All differences and standard deviations apply to the values of the variables. TDA Ball Mapper was performed on the normalised versions of each of the variables. Size is measured as the log of assets, and profitability as the return-on-equity. Diff. reports the difference between the means of the two groups being compared. Dist normalises the difference by the standard deviation of the variable in the whole dataset. All calculations performed using \textit{BallMapper} \citep{dlotko2019R}.}
\end{table}

Table \ref{tab:comp1} presents the results of comparisons involving five different sets of balls from the TDA Ball Mapper plot shown in Figure \ref{fig:multi1}. First consideration is given to the lower return ball number 26 which sits just below the common mass and is connected to darker blue, higher return balls. The nearest ball with the highest return is ball 11. Ball 11 has larger firms, the difference is negative, and these firms have greater investment. The firms in ball 11 also have a greater cashflow-to-price ratio. When normalising by the standard deviation it can be seen that the biggest difference between 26 and 11 comes from the level of investment, the second biggest from cashflow-to-price. Extending the comaprison to include the other high returning balls near ball 25, adding balls 1, 4 and 14 to ball 11 in the comparison, reveals a similar picture. However the difference in investment closes and leaves the cashflow-to-price ratio as the likely explanation for the difference. If a difference of two standard deviations is a benchmark then in the second case none of the variables meet the requirement. 

Next think about ball 25, a low return ball on the right hand extreme of the plot. This is connected to a high returning ball 10 and is on the opposite side of the plot from ball 12. Colouration by variables in panels (b) to (h) of Figure \ref{fig:multi1} showed much of the difference but it is still informative to have that as a relative magnitude. The third comparison in Table \ref{tab:comp1} shows that the main drivers of the difference are book-to-market, earnings-to-price and dividend yield. In the former two cases ball 25 has the highest values but in the final case it is ball 12 that offers the highest dividend yield. For two balls neighbouring each other to have such a difference is interesting, and challenges linearity assumptions. Comparing balls 25 and 10 reveals they are highly similar in almost every regard. Only the return to equity suggests a meaningful difference. The higher returning ball 25 has a lower return-to-equity than ball 10, seemingly contradicting the robust-minus-weak factor of \cite{fama2015five}. 

As a final comparison consider the top and bottom ends of the plot, balls 6 and 16. These both have very similar return levels but are shown to be far from connected. Figure \ref{fig:multi1} informed that earnings-to-price varied greatly between the two, but that it was profitability that had the greatest colour variation. When looking at this in numbers this is confirmed, as is a major difference in book-to-market ratio. That these contrive to produce similar returns despite so many differences says much about the importance of considering combinations of characteristics and the need to take great care when looking at regression results.

Five comparisons have been presented here but there are many more that can be done if the analyst so desires. From the colouration of the diagrams much can be learned but it is through drilling down into the data that the evidence to support the inference comes. 

\subsection{Regression Comparison}

In understanding the relationships between firm characteristics and stock returns it is typical to employ ordinary least squares (OLS) regressions. TDA Ball Mapper can contribute to the interpretation of the messages therefrom, informing on model fit and suggesting areas where better specification could be achieved. To illustrate this argument the following model is estimated:
\begin{equation}
    R_{i}=\alpha + \beta_1\ Size_i+ \beta_2\ BM_i + \beta_3\ ROE_i +\beta_4\ Invest_i + \beta_5\ EP_i + \beta_6\ Cash_i + \beta_7\ DY_i + \omega_i 
\label{eq:lm1}
\end{equation}
Data is taken from the time period to which the TDA Ball Mapper is applied, with the time subscripts omitted for clarity. Variables are as defined in the data section and $\omega_i$ is an identically, independently distributed error term with mean 0 and constant variance. Estimating equation \eqref{eq:lm1} for the five Junes discussed thus far produces Table \ref{tab:lm2}, where the figures in parentheses are the absolute values of the t-statistics for a test that each coefficient is equal to zero. Returns are expressed as percentages.

\begin{table}
    \begin{center}
        \caption{Monthly Regression Results ($\epsilon=0.40$)}
        \label{tab:lm2}
        \begin{tiny}
        \begin{tabular}{l c c c c c c c c c}
             \hline
             Year & Const. & Size & BM & ROE & Invest & EP & Cash & DY & Obs \\
             \hline
             June 1978 & 1.335 & -0.532*** & 0.037 & -0.743 & 0.495 & -1.147 & 0.005 & 0.142 & 2940 \\
             & (1.836) & (4.335) & (0.133) & (0.374) & (0.321) & (0.771) & (0.445) & (1.951)\\
             June 1988 & 2.217** & 0.588*** & -0.017 & 1.469* & 2.935** & 1.361 & -0.004 & -0.619 & 3625 \\
             & (3.279) & (4.512) & (0.039) & (2.533) & (3.015) & (1.428) & (0.788) & (0.583) & \\ 
             June 1998 & -0.792*** & 0.901*** & -0.060 & 1.676*** & -0.909 & 3.990** & 0.005 & 0.322** & 4954 \\
             & (11.96) & (8.345) & (0.148) & (4.158) & (1.068) & (3.208) & (1.821) & (2.947) &\\
             June 2008 & -5.286***& -0.732*** & -1.153 & -0.416 & 9.341*** & 6.815*** & 0.028*** & -0.468** & 3531 \\
             & (5.680) & (5.679 &(1.920) & (0.867) & (5.788) & (3.604) & (4.262) & (3.119) & \\
             June 2018 & -0.238 & -0.120 & -0.254 & 0.673** & 2.167 & 1.805* & 0.012* & -3.359 & 2837 \\
             & (3.024) & (1.162) & (0.810) & (2.765) & (0.911) & (2.336) & (2.095) & (0.386) & \\
             \hline
        \end{tabular}
        \end{tiny}
    \end{center}
\raggedright
\footnotesize{Notes: Estimation of equation \eqref{eq:lm1} performed on single month of data. Const. is the intercept ($\alpha$), size is the log market value of the firm, BM is the book-to-market ratio, ROE is the return on equity, Invest is the ratio of investment to assets, EP is the earnings per price, Cash is cashflow per price and DY is the dividend yield. Figures in parentheses are absolute values of the t-statistics. Significance given by * - 5\%, ** - 1\%, *** - 0.1\%.}
\end{table}

Residuals from these regressions may be readily used as a colouring variable for the TDA Ball Mapper plots as seen in Figure \ref{fig:multi1}.  Outliers in the parameter space have some of the highest residuals, but it is in the connected components that the interest lies. Residuals, panel (i) of Figure \ref{fig:multi1}, are at their largest in the net to the lower right of the central mass, this is a region where, despite similar small sizes, returns were highly variable. From the absolute residuals, panel (j) of Figure \ref{fig:multi1}, the notion of fitted error radiating out from the common mass is more strongly visible; balls close to the centre are seen with residuals of 0.05\% and more. Balls 3, 19 and 21 all fall into this category. Similar exercises with the other years confirms that model fit is always poorest in the areas where TDA Ball Mapper informs best about the combination of characteristics. Herein lies one of the many benefits of using approaches that do not pre-assume a particular relationship. 

Identifying areas of poor fit within the parameter space has a further advantage in that the researcher can consider which would be the optimum interaction terms to introduce to the regression model. Given that there would be significant correlation within the full set of interaction terms, being able to identify the best without relying on regression based optimisation techniques circumvents many of the fitting problems the multicolinearity would otherwise have created. This particular extension is left to future work.

\section{Annual Data}
\label{sec:annual}

Results presented thus far use monthly data such that each firm only appears once within the dataset. However, to go through each to deeply explore the messages from the data would be cumbersome. Annual plots avoid such weight of numbers but then embed the monthly variations in returns into one plot. For example the January anomaly would see smaller stocks giving very different returns and such may be lost in an annual point cloud. Considering annual data, this section uses the same $\epsilon=0.4$ employed in the monthly discussion to show how aggregation of data produces a further wealth of stories from within the data. A further note of caution is sounded on the issue of monthly variation and the response of the TDA Ball Mapper analyst documented. 

\begin{figure}
	\begin{center}
	    \caption{Common Anomalies TDA Ball Mapper Plots: 2018}
	    \label{fig:year1}
		\begin{tabular}{c c c}
			\multicolumn{3}{c}{\includegraphics[width=12cm,height=8.5cm]{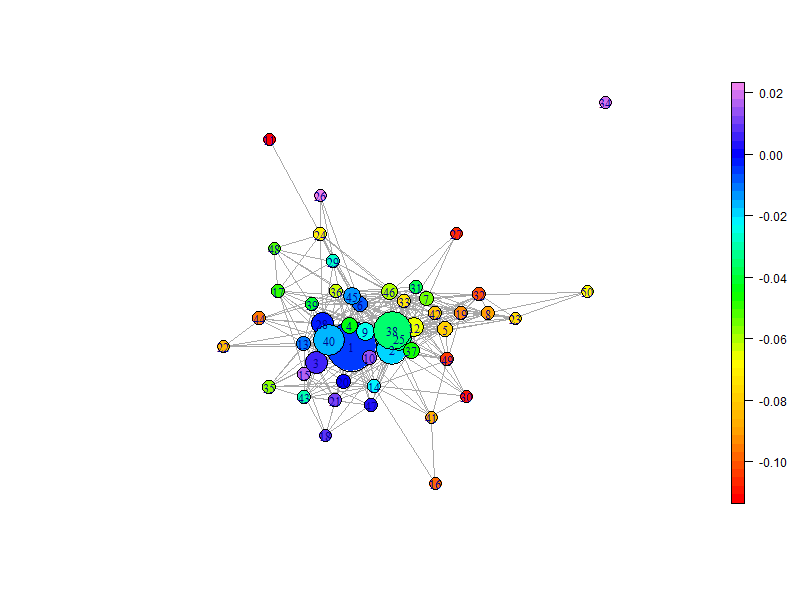}}\\
			\multicolumn{3}{c}{(a) Returns}\\
			\includegraphics[width=4.2cm]{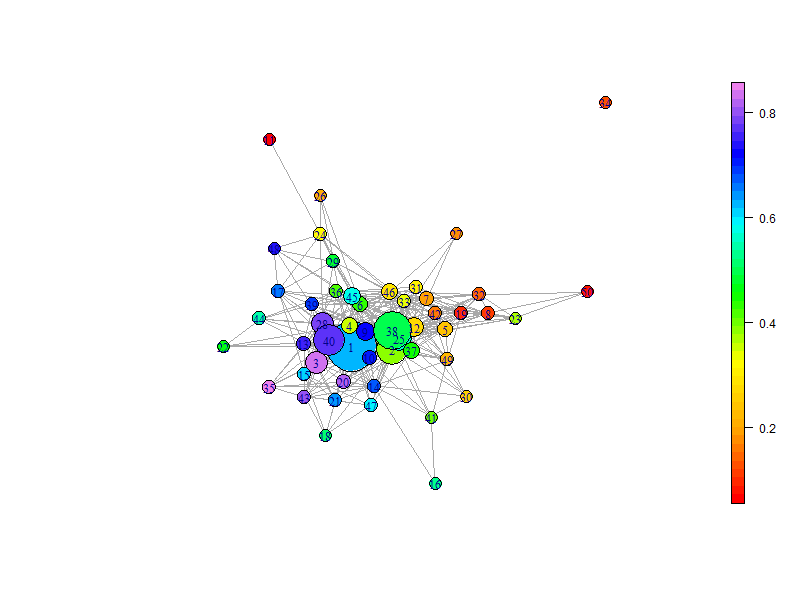}&
			\includegraphics[width=4.2cm]{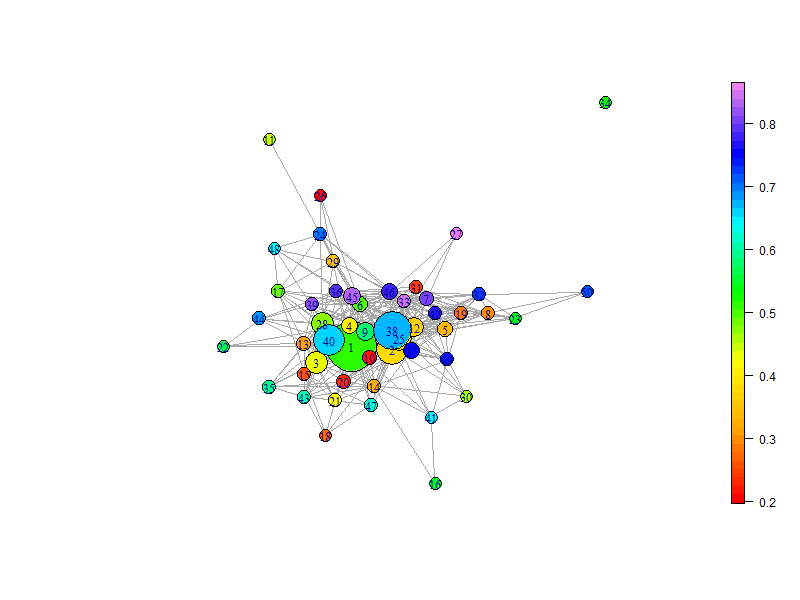}&
			\includegraphics[width=4.2cm]{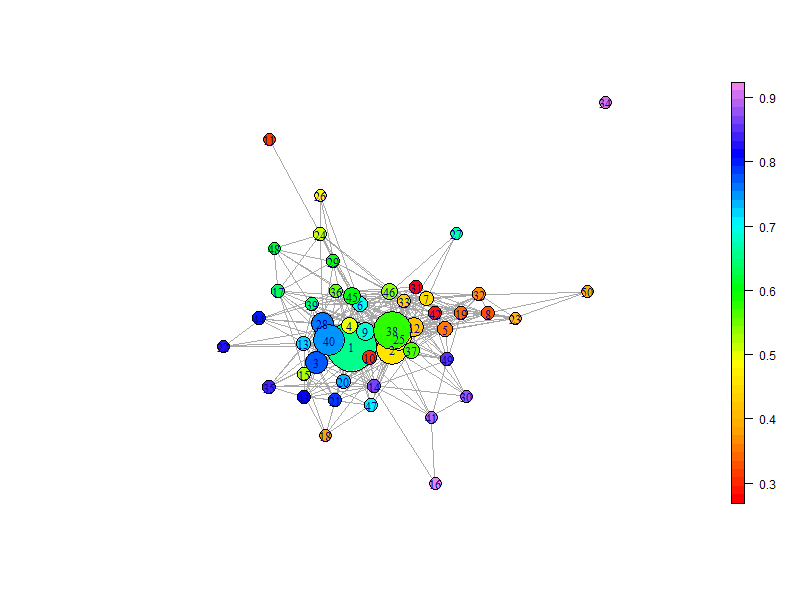}\\
			(b) Size & (c) Book-to-market & (d) Profitability \\
			\includegraphics[width=4.2cm]{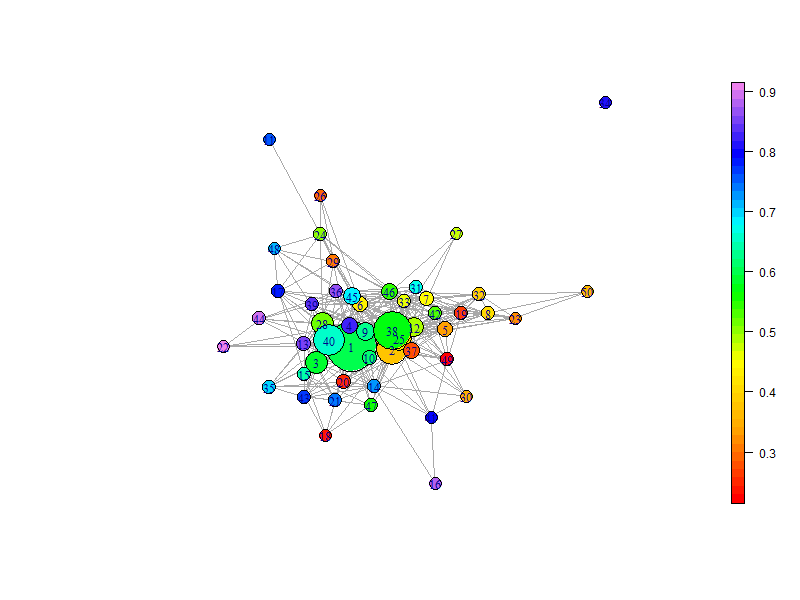}
			&
		    \includegraphics[width=4.2cm]{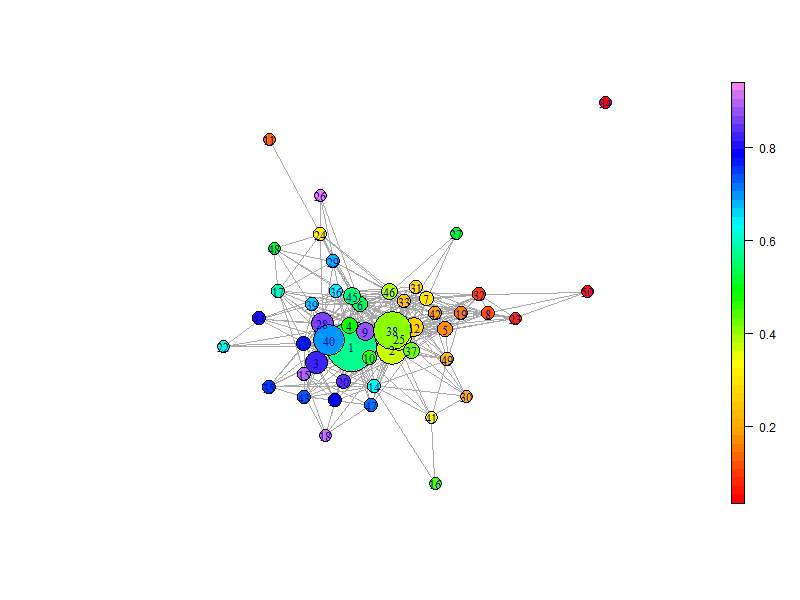}&
			\includegraphics[width=4.2cm]{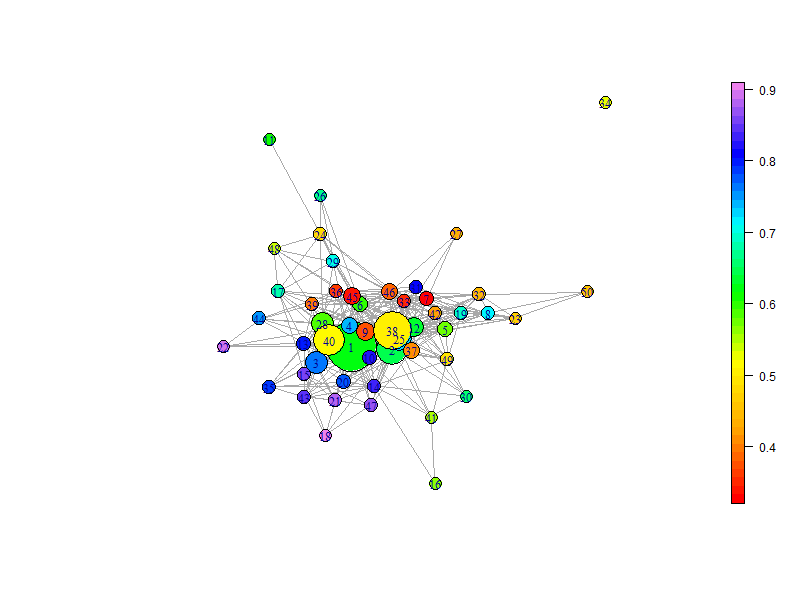}
			\\
			(e) Investment & (f) Earnings-to-price & (g) Cash-to-price \\

			\includegraphics[width=4.2cm]{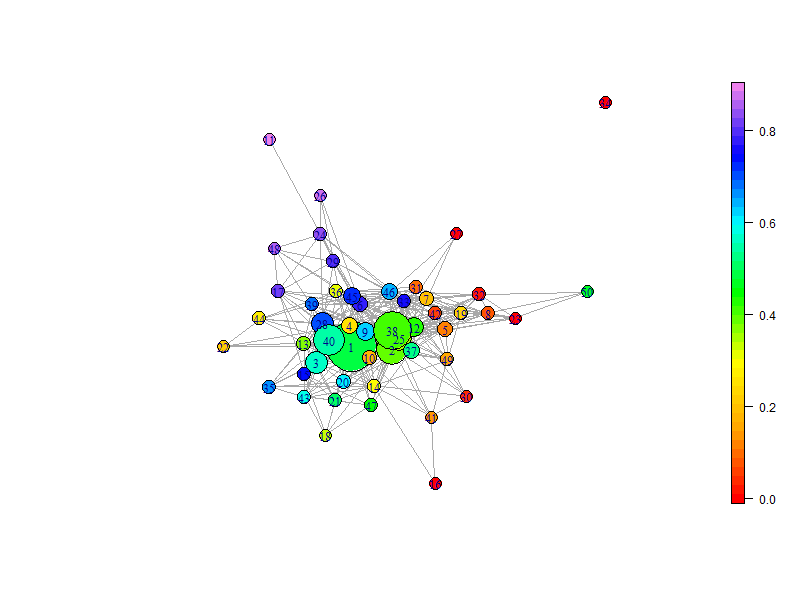}
			& \includegraphics[width=4.2cm]{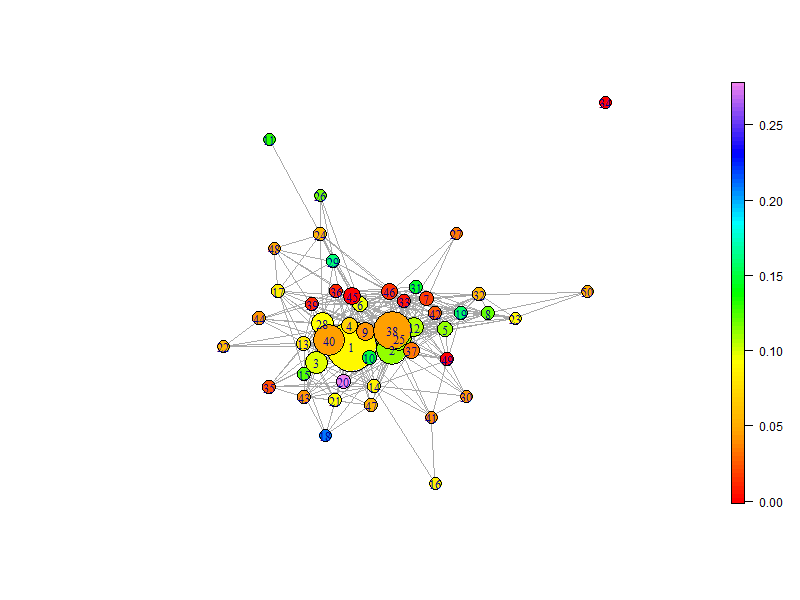}
			& \includegraphics[width=4.2cm]{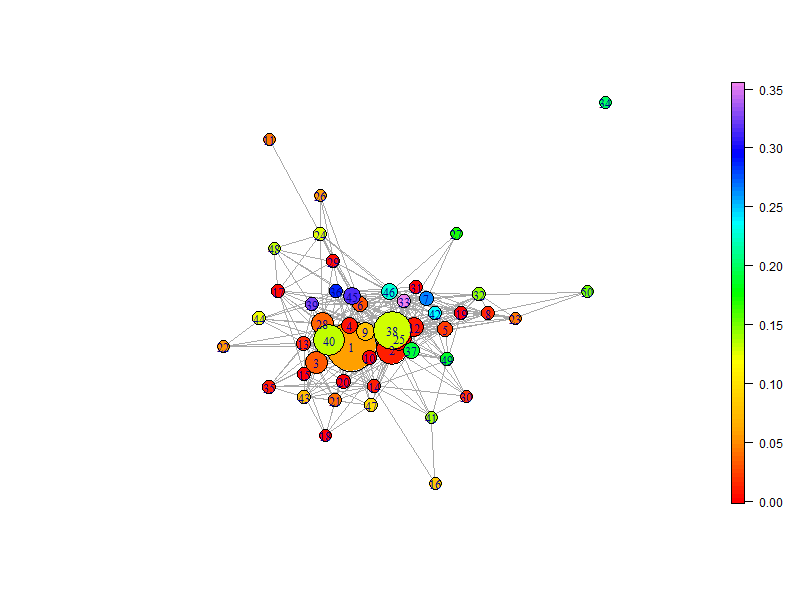}\\
			(h) Dividend yield & (i) March 2018 & (j) September 2018 \\
		\end{tabular}
	\end{center}
\raggedright
\footnotesize{Notes: TDA Ball Mapper plots generated using \textit{BallMapper} \citep{dlotko2019R} with radius $\epsilon=0.40$. Panel (a) coloured by average excess returns for the ball, panels (b) to (h) coloured by normalised values of the variables named below. In each case normalisation is to scale [0,1]. Residuals, panels (i) and (j) coloured by proportion of observations from that month. Colour versions available online.}
\end{figure}

Figure \ref{fig:year1} has many similarities with Figure \ref{fig:multi1}. Panel (a) shows that returns are highest in the left half of the common mass, and in the smaller balls below. Looking at the variable by variable plots below reveals that many of these are also displaying a similar pattern. Size in panel (b) is much more closely associated with the returns plot, having the same balls shaded in blue for high values. Profit in panel (d), investment in panel (e) and earnings-to-price in panel (f) also have a similar ordering amongst the main balls. For book-to-market and cash-to-price ball 1 has a lower value than 38, reversing the ranking seen in the returns. Moving into the extremes of the plots it is clear that once again the overall observed highs and lows are combinations of the various axis variables.

\begin{figure}
    \begin{center}
    \caption{TDA Ball Mapper Plots for Annual Returns}
    \label{fig:yearly2}
    \begin{tabular}{cc}
       \includegraphics[width=6.5cm]{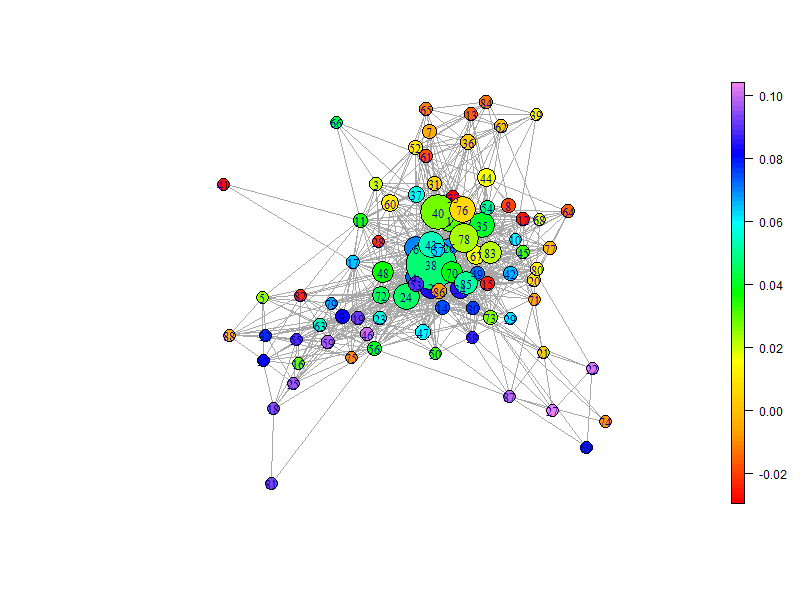}  &
       \includegraphics[width=6.5cm]{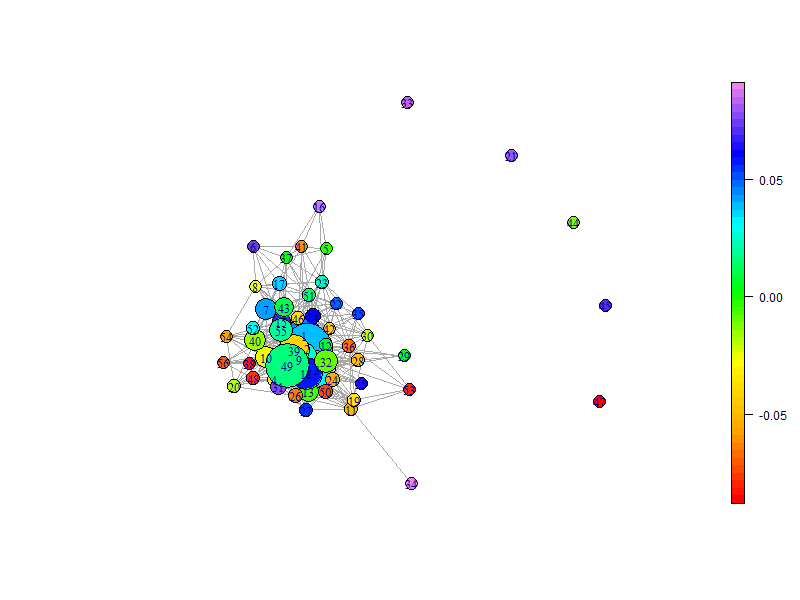}\\
       \includegraphics[width=6.5cm]{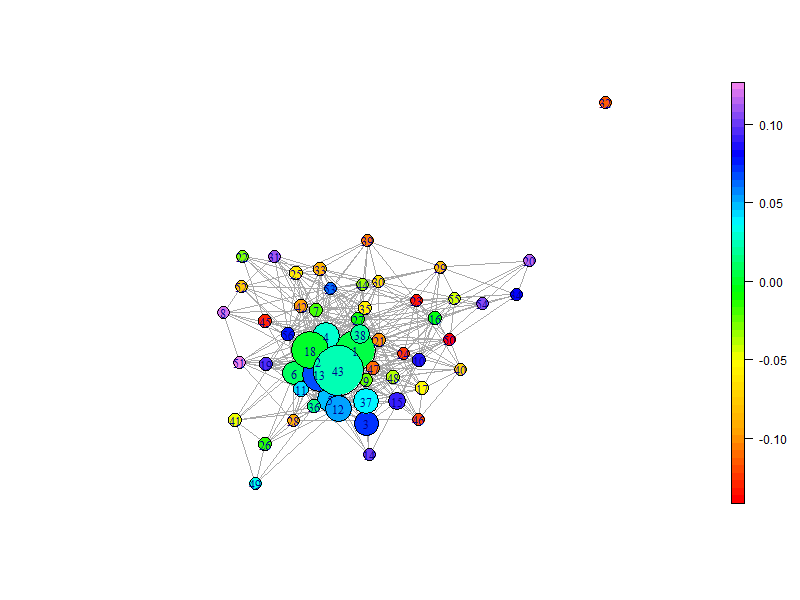}  & 
       \includegraphics[width=6.5cm]{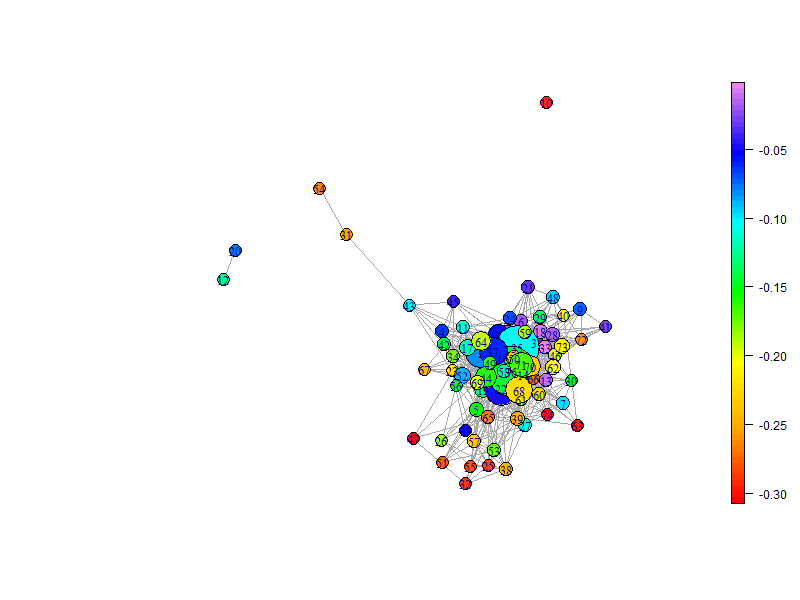}
    \end{tabular}
    \end{center}
\end{figure}

Looking at the other years very similar patterns emerge, with much more variation in returns than was seen in the monthly case. A potential explanation for this may be that the centre of the monthly plot is moving around through the year. If it follows that in some months the centre is closer to ball 1, and in others closer to ball 38, then it would follow that the months with high returns, and those with low returns would explain the variation between these two big balls. Using the replacement colouration functions of \textit{Ball Mapper} plots can be created that use a dummy for each month as the colouring variable. These plots, of which only two are included here for brevity, reveal that the months do move around the space. The contrast between March in panel (i) of Figure \ref{fig:year1} and September in panel (j) shows well how different months concentrate in different parts of the annual point cloud. Thus month explains the shape seen, but not the colouration. March observations make up a higher proportion of the bottom left of the diagram with some spread through the top left. September, by contrast is more represented in the areas immediately above the central group with a much lower percentage of those in ball 1 and 2 than March. Using the individual variable colouration links the balls with more March observations to higher returns, whilst September tallies more with higher dividend yield. 

Through a brief exploration of the annual data a case for the study of individual months is made. Everything that can be done from the monthly data may be done to this annual data, but obtaining usable inference is best achieved from monthly plots.

\section{Discussion}
\label{sec:discuss}

TDA Ball Mapper is a valuable data visualisation and exploration tool that has been employed here to demonstrate non-linearities and patterns within the data that linear modelling of the firm characteristic-outcome relationship risks missing. In beginning an exploration the analyst seeks more understanding of what lies within the data they hold; this is delivered. However, TDA Ball Mapper is not the only way to summarise such data; alternatives such as k-means clustering can be quickly used to inform about groupings within the characteristic space. Likewise, the wealth of information that comes out of the TDA Ball Mapper requires a great deal of human involvement to interpret. To the former concern this section offers a discussion of results from clustering analyses. To the latter a selection of next steps for the research are explored. Resulting is a deeper appreciation of how TDA Ball Mapper can contribute to the development of stock return modelling.

\subsection{Clustering and TDA Ball Mapper}

TDA Ball Mapper partitions the data into a collection of balls and joins with edges the neighbouring ones. Consequently well separated clusters will appear as different connected components of the TDA Ball Mapper graph. In this way it is analogous to the clustering algorithms regularly applied within data analysis. However, unlike such algorithms the aim of TDA is not to achieve any form of even distribution across the space, TDA does not have any weighting on characteristics and, owing to the overlapping of balls in the net, TDA allows points to appear in multiple groups in a way most clustering algorithms do not. To illustrate these differences within the common anomaly dataset a comparison with k-means clustering \citep{hartigan1979algorithm} is undertaken. In setting up the methodology it was noted that TDA Ball Mapper is not a clustering tool. This section undergirds that point with an example from the data.

The k-means clustering seeks to partition the data into $k$ groups such that within the clusters the variation is minimised. For a given $k$ the sets are constructed in the way that the sum of the squared difference between each datapoint and the centre of the cluster is minimized. This desire to minimise variation would be analogous to TDA Ball Mapper if points were evenly spread in all dimensions, but as this is seldom true an inevitable difference occurs. A single parameter, the number of clusters, must be selected and, like the choice of $\epsilon$ the level chosen is a trade off between detailed small groupings and the low noise ease of interpretation afforded by larger groups. Optimal cluster numbers are set using R package \textit{mclust} \citep{mclust2016r}. 

\begin{table}
    \begin{center}
        \caption{Comparison of k-means and TDA Ball Mapper}
        \label{tab:comp}
        \begin{tabular}{l c c c c c c c c}
             Period & \multicolumn{3}{l}{k-means (Optimal)} & \multicolumn{3}{l}{TDA Ball Mapper ($\epsilon=0.40$)} & \multicolumn{2}{l}{k-means ($m$)}\\
             & Number & Min & Max & Number ($m$) & Min & Max & Min & Max \\
             \hline
             June 1978 & 19 & 32 & 281 & 33 & 1 & 1415 & 16 & 170  \\
             June 1988 & 37 & 9 & 361 & 27 & 1 & 1544 & 8 & 383\\
             June 1998 & 39 & 9 & 336 & 28 & 1 & 3181 & 8 & 515\\
             June 2008 & 35 & 9 & 335 & 31 & 1 & 2755 & 9 & 392\\
             June 2018 & 35 & 6 & 241 & 27 & 1 & 2438 & 4 & 321 \\
        \end{tabular}
    \end{center}
\raggedright
\footnotesize{Notes: Comparison table constructed for five months. k-means (Optimal) reports from the optimum number of clusters selected by \textit{mclust} \citep{mclust2016r}. $m$ is number of balls created by TDA Mapper. Min and Max denote the size of the smallest and largest group. k-means ($m$) is then k-means clustering performed with $m$ clusters. Where TDA Ball Mapper assigns a point to multiple balls the lowest ball number is used. Min and Max report the size of the smallest and largest clusters in that year.}
\end{table}

Comparison of the methodologies in Table \ref{tab:comp} reveals that k-means commonly selects more clusters than the $\epsilon=0.40$ TDA Ball Mapper; variation of $\epsilon$ would create more clusters however. Numbers of observations in each grouping is more tricky for TDA Ball Mapper since many points are in multiple balls; this follows from the need to have points in the intersection. To facilitate comparison such points are allocated to the lowest ball number, but other tie-break rules produce similar inference on the size of the largest balls. A better demonstration is seen when the number of groups generated by the two approaches is the same. This is shown in final block of Table \ref{tab:comp}. An immediate observation is that TDA Ball Mapper produces groups with just one observation, but the k-means approach never generates a group so small. Similarly the largest groupings in the TDA Ball Mapper set are notably bigger than those produced in k-means, often more than 5 times larger. Between the two extremes the group sizes in k-means are much more evenly distributed than the ball sizes are in the TDA Ball Mapper case.  

TDA Ball Mapper thus preserves the topographical information from all points, even where this means single observation groups. Determination of whether such individual cases merit further investigation is left to the user. As the aim in this paper is to compare regions of the point cloud it must be possible to make statement like ``group X has return Y, whilst neighbouring group V has return W`` and hence assign the differences between Y and W to differences in X and V. The topological faithfulness of TDA Ball Mapper means cross examination of characteristic values is straightforward and, critically, directly linked to the outcome. Introducing a third group with characteristics S, and returns T, becomes seemless because the relative positions of X, V and S are easily verified. Clustering algorithms do not offer such; comparing clusters needs much more analysis of the centroid coordinates and need reappraisal every time a new cluster is added to the evaluation. It is stressed again TDA Ball Mapper is not purposed as a clustering algorithm and this section serves to empirically demonstrate the implications thereof.

 \subsection{Implementations and Next Steps}

Investors may find value in the outputs of TDA Ball Mapper where they identify balls with high returns amongst characteristic combinations that yield the low returns standard models would be. Those armed with simple models would ignore firms that have a particular combination of ratios that TDA Ball Mapper consistently identifies as being worth investing based upon. This seven variable set has already shown that some combinations bring high returns even where none of the individual axis variables would normally have said to expect high returns. In identifying stocks of interest the approach is also reminding on the danger of following simple rules for share selection. 

Faced with the outcome of a TDA Ball Mapper plot an investor may identify where the returns are high and interrogate to find what combination of ratios produced that outcome. They may further conduct pairwise comparisons against neighbouring balls to know more of what specifically was linked to the good performance. Understanding what is likely to then happen to a firms ratios in future periods can say something to the returns the investor is likely to get by taking out a position on that stock. Evidence presented in this paper has shown consistency across months that may permit such strategy. Trading strategies are beyond the scope of this paper however.

TDA Ball Mapper is an exploratory data analysis tool which is informed by the topology of the underlying dataset. Free from any pre-conceptions about expected relationships, or variable interpretations, it constructs an abstract representation of the data to hand. Because it is a bottom-up approach analysis of the output is needed to make sense of what is being demonstrated. To this area there is scope for significant further contributions across the inter-disciplinary sphere. This paper has shown similarities persisting across time but has not sought to develop a measure to confirm that. Interesting pairwise comparisons were highlighted but no attempt was made to formalise identification of which pairs to compare. A role here is envisaged for machine learning alogtithms that seek out sets of interesting patterns from TDA Ball Mapper graphs. The potential agenda in this direction is rich.

\section{Conclusions}
\label{sec:conclude}

This simple exposition has sought to introduce the TDA Ball Mapper algorithm of \cite{dlotko2019ball} as a means of understanding the relationships, and interdependencies, of firm characteristics in the determination of stock returns. Using a two axis example it was shown that links are non-linear in the way that existing literature assumes. Extending to a seven dimensional set based upon commonly studied stock return determinants the value of the TDA approach in unpacking the correlations between financial ratios and stock returns was again demonstrated. Whilst variation over time is natural the broad picture remains constant throughout the forty year sample. Resulting is a deeper appreciation of what this extensive period of financial data relays about the market.

As presented TDA Ball Mapper is reliant on a single parameter, the ball radius $\epsilon$,  which determines how refined the analysis will be. To date there is no definition of what the optimal level should be. At one hand it is a weakness which requires further research to develop an appropriate metric that can remove the reliance on analyst integrity. On the other hand we need to acknowledge the existence of data sets, most notably fractals, that possess different features on different scales. For them, there is no canonical way of choosing the resolution and in general, it is not possible to determine if any given data set has such a feature.  Decisions on how much attention to pay to single observation balls are also challenging in the context of other methodologies. With thoughtful research design, and due consideration of the outcomes, the power of TDA Ball Mapper can be exploited with no unnecessary risk to the quality of the outcomes.

From an investment perspective the identification of stocks which do not behave according to their type is clearly of interest. Such stocks stand out clearly in the mapping, especially at finer resolutions, and demonstrate the dangers of using trading models premised on aggregate relationships. Development of trading models using TDA Ball Mapper output is a key avenue for future research.  As TDA Ball Mapper uses a stand-alone colouring function it would be trivial to replace current return with return for the following month, or any other information that the investor would like to see. This is built in to the \textit{BallMapper} pacakage \citep{dlotko2019R}. 

This paper offers a new understanding of the link between established firm financials and stock returns. Using a robust methodology that is built from the data, and faithfully respects the topology thereof, questions are raised about the results that have emerged from pre-assumed monotonic relationships. Early promise for investors has been identified and it is for subsequent work to build on that strong foundation. Future research on the fuller set of characteristics offers potential, as does the ability of TDA Ball Mapper to work in data reduction, in dynamic settings and in conjunction with machine learning. A rich seam of exploration awaits. For investors, analysts and those seeking a deeper understanding of markets an invaluable tool to undergird effective organisation is provided. The full power of the tool is there to be exploited.

\bibliographystyle{apalike}
\bibliography{interactcadsample}
\end{document}